\documentclass[aps,showpacs,nofootinbib]{revtex4-1}
\usepackage{amsmath}
\usepackage{amsfonts}
\usepackage{amssymb,amscd,amsbsy}
\usepackage{epsfig}
\usepackage{color}

\usepackage{mathtools}

\newcommand{\bea}{\begin{eqnarray}}
\newcommand{\eea}{\end{eqnarray}}

\newcommand{\vect}[1]{\mathbf{#1}}

\newcommand{\req}{\rho_{\rm eq}}

\newcommand{\tb}{\textcolor{black}}
\newcommand{\tm}{\textcolor{black}}
\newcommand{\tr}{\textcolor{black}}
\newcommand{\kt}{k_{\rm B}T}

\newcommand{\cref}{c^{(2)}_{\rm ref}}

\newcommand{\imag}{{\rm i}}

\begin{document}
\title{Description of hard sphere crystals and crystal--fluid interfaces:
a critical comparison between density functional approaches and a phase field crystal model}

\author{M. Oettel$^{1,2}$, S. Dorosz$^3$, M. Berghoff$^4$, B. Nestler$^4$ and T. Schilling$^3$ }
\affiliation{ $^1$ Johannes Gutenberg--Universit\"at Mainz, Institut f{\"ur} Physik,
  WA 331, D--55099 Mainz, Germany \\
$^2$ Institut f\"ur Theoretische Physik II, Heinrich-Heine-Universit\"at D\"usseldorf, D--40225 D\"usseldorf, Germany  \\
$^3$ Universit\'e du Luxembourg, Theory of Soft Condensed Matter, L-1511 Luxembourg, Luxembourg \\
$^4$ Institute of Applied Materials, Karlsruhe Institute of Technology, D--76131 Karlsruhe, Germany 
}


\begin{abstract}
\tb{ In materials science the phase field \tr{crystal} approach  \tr{has become popular} to model 
crystallization processes.  Phase field \tr{crystal} models are in essence 
Landau--Ginzburg--type models, which should be derivable from the underlying  
microscopic description of the system in question. We present a study on 
classical density functional theory in three stages of approximation \tr{leading to a specific 
phase field crystal model}, and we discuss the 
limits of applicability of the models that result from these approximations.
As a test system we have chosen the three--dimensional suspension of 
monodisperse hard spheres. \\
The levels of density functional theory that we discuss are 
fundamental measure theory, a second--order Taylor expansion thereof, and \tr{a minimal} 
phase--field crystal model. We have computed coexistence densities, vacancy 
concentrations in the crystalline phase, interfacial tensions and interfacial 
order parameter profiles, and we compare these quantities to simulation 
results. We also suggest a procedure to fit the free parameters of the 
phase field crystal model.\\
In brief, we conclude that fundamental measure theory is very accurate and 
can serve as a benchmark for the other theories. Taylor expansion strongly 
affects free energies, surface tensions and vacancy concentrations. 
Furthermore it is \tr{phenomenologically} misleading to interpret the phase field crystal model as 
stemming directly from Taylor--expanded density functional theory.}
\end{abstract}


\maketitle

\section{Introduction}
In materials science, the modelling of dynamic processes involving
the growth of solid phases in melts or in the environment of another solid phase
has been advanced using phase field models in the past years 
\cite{Emm08, Hoy03}.
Here, the phase field $\varphi$
is associated with an order parameter field \tb{ that distinguishes between} 
a solid and a liquid
phase, and it is usually coupled with a density or concentration field $\varrho$
for that phase. The dynamics of $\varrho$ is conserved, following the equation
\bea
 \label{eq:ddft1}
 \frac{\partial\varrho}{\partial t} &=&  \nabla \cdot \left( \Gamma_\varrho \vect j_\varrho\right) \; \\
 \label{eq:ddft2}
        \vect j_\varrho(\vect r, t) & = &  \nabla \frac{\delta {\cal F}[\varrho,\varphi]}{\delta \varrho(\vect r,t)}.
\eea
Here, the density current $\vect j_\varrho$ is the gradient of a chemical potential function
which is assumed to be derivable from a free energy functional ${\cal F}$. $\Gamma_\varrho$
is a mobility. In contrast to this,
there is no conservation law for the order parameter and thus one can generically assume
a nonconserved dynamic evolution of $\varphi$ of the form
\bea
  \frac{\partial\varphi}{\partial t} & =& \tilde\Gamma_\varphi \frac{\delta {\cal F}[\varrho,\varphi]}{\delta \varphi(\vect r)}.
\eea
\tb{ In order to briefly explain the approach,}
we consider a one--component system able to form one fluid and one solid phase.
The simplest free energy functional which gives us phase coexistence associated with
smoothly varying profiles for $\varrho$ and $\varphi$ across the phase boundary follows from a gradient expansion
in the specific truncation
\bea
 \label{eq:phase_field_f}
   {\cal F}[\varrho,\varphi] &=& \frac{F_0}{\kt} \int d^3 r \left( c_2 (\nabla\varphi)^2 +
   f(\varrho,\varphi) \right) \;.
\eea
Here it is assumed that any inhomogeneity costs free energy through the gradient term in the
order parameter field. Since \tb{ there is no} corresponding gradient term in 
the density, the
free energy penalty corresponding to a change in the density field must be small, and this
appears to be only possible if $\varrho$ is suitably coarse--grained from the microscopic
density field. Consequently, the variations in the microscopic density field  relevant
for the free energy must be contained in the phase field $\varphi$ which in turn should
be derivable from the microscopic density through another coarse--graining procedure.
We will return to that point below. The potential function $f(\varrho,\varphi)$ contains
a double--well type expression with minima at $\varphi=-1$ (fluid) and $\varphi=1$ (solid), modified such that
minimization with respect to $\varrho$ gives the input fluid and crystal coexistence densities
$\varrho_{\rm fl}(T)$ and $\varrho_{\rm cr}(T)$ which in general depend on the temperature
$T$. A possible form is \cite{Hoy03}
\bea
 f(\varrho,\varphi) &=& g_1(\varphi) + \frac{1}{2}(1+g_2(\varphi)) f_{\rm cr}(\varrho) +
                        \frac{1}{2}(1-g_2(\varphi)) f_{\rm fl}(\varrho) \; ,\\
 \text{with}  &&  g_1(\varphi) = (\varphi+1)^2(\varphi-1)^2 \;, \\
 \text{and}   &&  g_2(\varphi=\pm 1) =\pm 1\;, \quad g_2'(\varphi=\pm 1) = 0 \;.
\eea
Thus, the \tb{ phase field} approach is nothing but a slighly rewritten Landau--Ginzburg model for
the fluid--solid phase transition. The formulation accomodates an empirical free energy density
for the fluid phase ($f_{\rm fl}(\varrho)$) and the crystal phase ($f_{\rm cr}(\varrho)$)
yielding the required input coexistence densities.
In the form of Eq.~(\ref{eq:phase_field_f}), the free energy contains the parameters
$F_0$ related to \tm{is} the free energy scale in units of the thermal energy $\kt$
(should be adjusted to the bulk free energy difference of solid and liquid) and the constant
$c_2$ which can be fixed through the value of the liquid--solid surface tension.

In such a way, nucleation and growth in simple systems can be addressed without resolving the details of the free energy for inhomogeneous systems.
For the widely used hard sphere reference system (which will be examined more in detail
in this work), this has been done in e.g.
Ref.~\cite{Tot09}\tm{.}

Such a minimal Landau--Ginzburg description can be extended to more complex systems.
For each new material component of the sytem, one needs to introduce a corresponding density field,
and new phase fields for the fluid and solid phases,
even when the solid phases just differ by their crystalline orientation. Thus
the number of free energy and surface tension parameters quickly grows
when the complexity of the system is increased. \tb{ Even} for the 
one--component system, empirical information on the anisotropy of the 
surface tension for different crystal faces in equilibrium with the 
fluid \tb{ needs to be taken into account to set the parameters}. 
Hence an important question is
whether the phase field itself can be consistently treated \tb{ in terms of} 
the density field \tb{ which stems} from a microscopic foundation.

In this paper, we investigate three formulations (or approximations) to classical density
functional theory which deal with the microscopic particle density field and thus, in principle,
constitute the underlying theoretical framework from which a consistent phase field crystal
description should arise. In our explicit calculations, we examine the coexistence properties,
surface tensions and interface density modes in the hard sphere system. Understanding surface tensions
for different interface orientations and the associated surface structure are important
prerequisites for further studies.
There are three main reasons to choose the hard sphere system for method comparison: (1) availability
of very precise density functionals (fundamental measure theory, our first formulation).
(2) empirical evidence that the crystal--liquid surface tensions of fcc(hcp) forming metals are largely
of entropic origin and thus the packing of impenetrable cores \tm{plays} an important role for
the surface structure of these metals \cite{Lai01} and (3) the athermal nature of the hard sphere system
which reduces the parameters for describing the coexistence to just the pair of coexistence densities for the liquid
and solid phase. -- Our second formulation is Taylor--expanded density functional theory which
neglects the density fluctuations with respect to a reference density beyond second order in the free energy formulation.
It constitutes already a drastic approximation in density functional theory, nevertheless it is occasionally
depicted in the literature as ``the" density functional theory from which the third formulation investigated here,
the phase field crystal model of the simplest type, can be derived.
In short, the phase field crystal model can be viewed as a local expansion in density fluctuations and in their gradients
of Taylor--expanded density functional theory.
 The model (see below for its description and references) has lately come to some prominence in the materials science
community mainly for the reason alluded to a microscopic foundation of phase field descriptions, (see
Ref.~\cite{Gol05} for a systematic attempt in that direction).

The paper is structured as follows. Sec.~\ref{sec:dft_intro} introduces briefly the foundations of classical
density functional theory and describes the formal approximation steps leading to Taylor--expanded density
functionals  and phase field crystal models. In Sec.~\ref{sec:results}, the explicit functionals are given and
they are applied to the hard sphere system. In calculating surface tensions and the interface structure,
no further approximations are made in order to avoid uncertainties in interpreting the results. We concisely discuss
the problem of parameter fixing for the phase field crystal model. Sec.~\ref{sec:summary} contains
our summary and conclusions.

\section{Density functional theory and phase field crystal models}

\label{sec:dft_intro}

As discussed before, the phase and the density field in the phase field formulation
should be both obtainable through
a suitable coarse graining of the microscopic density field.
Thus one may be tempted to forego the artificial distinction between (coarse--grained)
phase and (coarse--grained) density field entirely in favor of the microscopic density
$\rho(\vect r)$. Indeed, in equilibrium the basic theorems of density functional theory
assure us that there is a unique free energy functional of the one--particle
density field $\rho(\vect r)$,
\bea
  {\cal F}[\rho] & = &{\cal F}^{\rm id}[\rho] + {\cal F}^{\rm ex}[\rho] \;, \\
  \label{eq:fid}
  \text{with } \beta  {\cal F}^{\rm id}[\rho] &=& \int d^3 r \rho(\vect r)\left( \ln(\rho(\vect r)\Lambda^3) -1 \right)
\eea
which can be split into the exactly known ideal gas part ${\cal F}^{\rm id}$ ($\Lambda$ is the de--Broglie wavelength,
$\beta=1/(\kt)$ is the inverse temperature)
and a generally unknown
excess part ${\cal F}^{\rm ex}$. The equilibrium density $\req$ in the presence of an external
(one--particle) potential $V^{\rm ext}(\vect r)$ is then given by
\bea
 \label{eq:minimizingF}
  \left.\frac{\delta {\cal F}[\rho]}{\delta\rho(\vect r)}\right|_{\rho=\req} = \mu - V^{\rm ext}(\vect r)\;,
\eea
where $\mu$ is the imposed chemical potential (e.g. by requiring a certain bulk density far away from the
region where the external potential acts). For diffusive dynamics, the time evolution of this microscopic one--particle
density field obeys the type of dynamics as in Eqs.~(\ref{eq:ddft1}) and (\ref{eq:ddft2}):
\bea
 \label{eq:ddft3}
 \frac{\partial\varrho}{\partial t} &=&  \Gamma \nabla \cdot \left(  \rho(\vect r, t)
  \nabla\left[ \frac{\delta {\cal F}[\rho]}{\delta \rho(\vect r, t) } + V^{\rm ext}(\vect r, t)  \right]\right) \;
\eea
To show this, one needs the possibly severe approximation that the time--dependent density--density correlation function
can be approximated by the corresponding equilibrium object which in turn is obtainable from
the equilibrium density functional ${\cal F}[\rho]$ \cite{Arc04}. Note that the density field $\rho(\vect r,t)$ is an
ensemble--averaged quantity with no coarse--graining in space and time and there is no noise term in
Eq.~(\ref{eq:ddft3}).

\subsection{Functional Taylor expansion}

\label{subsec:taylor}

Since the excess free energy functional ${\cal F}^{\rm ex}$ is unknown in general, many practical applications of DFT
have started from an expansion of ${\cal F}^{\rm ex}$ around a background reference density profile $\rho_0(\vect r)$ which,
in general, can depend on the position:
\bea
 \label{eq:f_hnc}
 \beta {\cal F}^{\rm ex} = \beta F^{\rm ex}_0[\rho_0] - \int d^3 r c^{(1)}(\vect r;\rho_0)\Delta\rho(\vect r) - \frac{1}{2} \int d^3 r d^3 r'
    c^{(2)}(\vect r, \vect r';\rho_0) \Delta\rho(\vect r)\Delta\rho(\vect r') + \dots
\eea
Here, $F^{\rm ex}_0[\rho_0]$ is the excess free energy pertaining to the background profile,
$\Delta\rho(\vect r) = \rho(\vect r)-\rho_0(\vect r)$ and $c^{(1)}$ and $c^{(2)}$ are the first two members in the hierarchy
of direct correlation functions $c^{(n)}$, defined by
\bea
 \label{eq:cn_def}
 c^{(n)}(\vect r_1,\dots,\vect r_n;\rho_0) = -\beta \left.
   \frac{\delta^{(n)} {\cal F}^{\rm ex}}{\delta \rho(\vect r_1)\dots \delta \rho(\vect r_n)}
   \right|_{\rho=\rho_0(\vect r)} \;.
\eea
In most practical applications, $\rho_0\equiv const.$ is taken to be a reference bulk density in which case
$-c^{(1)} = \beta \mu^{\rm ex}= \beta\mu-\log(\rho_0\Lambda^3)$ and $c^{(2)}(\vect r-\vect r';\rho_0)$ depends only on
one position. To evaluate the functional in Eq.~(\ref{eq:f_hnc}), the correlation function $c^{(2)}$ has to be determined as an external input, provided e.g.\ by integral equation theory or by simple approximations of RPA type \cite{Eva79}.

It is perhaps somewhat surprising that the functional in Eq.~(\ref{eq:f_hnc}) (with $\rho_0$ being a bulk density)
is capable of describing fluid--solid coexistence. This has been shown first in Ref.~\cite{Ram79} for the case of hard spheres
(for an {fcc} crystal structure)
with $c^{(2)}$ taken to be the analytically known solution of the Percus--Yevick closure to the integral equations.
After all, the direct correlation function in the {\em solid} phase should be very distinct from the one in
the {\em liquid} phase, as can be inferred from the definition in Eq.~(\ref{eq:cn_def}). Consequently
the expansion should only hold for modest departures from the reference bulk density which is not the case
when comparing the density distribution in the crystal and the liquid, owing to the occurence
of sharply peaked crystal density profiles. However,
the {fcc} crystal density profile probes the Fourier transform $\tilde c^{(2)}(k;\rho)$  at discrete points in $k$--space
(the reciprocal lattice vectors (RLV)) which include the points where the structure factor has its maxima.
Furthermore, the $\vect k$--vectors of the RLV are distributed fairly isotropically (see also a more detailed
discussion on that in Ref.~\cite{Oet10}).

With suitable input for $c^{(2)}$, the Taylor expanded functional in Eq.~(\ref{eq:f_hnc}) is also capable of describing
the fluid--bcc transition (relevant for the description of e.g. iron). See Ref.~\cite{Jaa09} for a recent study. 
However, the numerical results for free energies and also for crystal--liquid surface tensions
obtained in such studies do not compare well with simulation results, e.g., the surface tensions from Ref.~\cite{Jaa09} deviate by a factor of 2. Thus, the Taylor--expanded functional appears
to be merely a suitable qualitative tool to explore basic features of dense liquids in the vicinity of the solid or glass
transition (see e.g. \cite{Hop10, Sin11}).

\subsection{The phase field crystal model}

\label{subsec:pfc}

\begin{figure}
 \epsfig{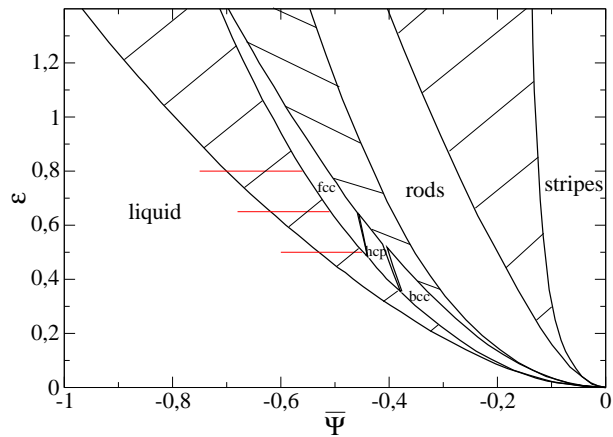}
 \caption{(color online) Phase diagram of the PFC free energy (\ref{eq:fpfc}). Data are taken from Ref.~\cite{Jaa10}. The horizontal
  lines mark the values $\epsilon=0.5$, 0.65 and 0.8 for which explicit results are discussed (see below).
 }
 \label{fig:phasediag_pfc}
\end{figure}

The Taylor expanded functional in Eq.~(\ref{eq:f_hnc}) is nonlocal in the densities. Through an additional approximation (gradient
expansion) it can be cast into a local form. We consider again a constant reference density $\rho_0$ and the following
power expansion of the Fourier transform of the direct correlation function:
\bea
   \tilde c^{(2)} (k;\rho) = -c_0 +c_2\; k^2 - c_4\;k^4 \dots
\eea
Using this, the Taylor--expanded functional becomes
\bea
   \beta {\cal F}^{\rm ex}_{\rm loc} = \beta F^{\rm ex}_0(\rho_0) + \beta \mu^{\rm ex} \int d^3 r \Delta \rho(\vect r) +
             \frac{1}{2} \int d^3 r \Delta\rho(\vect r)\left( c_0 + c_2\nabla^2 + c_4 \nabla^4 \dots   \right)\Delta\rho(\vect r)
    + \dots \nonumber \\
\eea
We observe that the excess free energy density contains local terms up to order 2 in $\Delta\rho$ and up to order 4 in $\nabla(\Delta\rho)$.
The total free energy contains additionally the ideal gas term, ${\cal F}^{\rm id}[\rho]$ from Eq.~(\ref{eq:fid}).
One may expand also this term in $\Delta\rho$ in order to obtain a consistently power--expanded free energy density.
It has been customary in the literature to introduce the dimensionless density difference as an order parameter:
\bea
 \phi(\vect r) = \frac{\rho(\vect r) - \rho_0}{\rho_0}\;.
\eea
In terms of $\phi$, the power--expanded total free energy up to order 4 in $\phi$ and $\nabla\phi$ reads
\bea
 \beta {\cal F} &\approx & \beta F_0(\rho_0) + \int d^3 r \left( \rho_0 \beta\mu \phi(\vect r) +
       \frac{a_{\rm id} \phi^2(\vect r)}{2} -\frac{b_{\rm id} \phi^3(\vect r)}{3} +
       \frac{g_{\rm id} \phi^4(\vect r)}{4}\right) + \nonumber   \\
   \label{eq:pfc1}
     & & \frac{\rho_0^2}{2} \int d^3 r \phi(\vect r)\left( c_0 + c_2\nabla^2 + c_4 \nabla^4  \right)\phi(\vect r) \;.
\eea
The coefficients $a_{\rm id}=\rho_0$, $b_{\rm id}=\rho_0/2$, and $g_{\rm id}=\rho_0/3$ stem from the expansion
of the ideal gas part of the free energy.
The reference free energy $F_0$ contains the ideal gas part by $F_0(\rho_0) =  F^{\rm ex}_0(\rho_0) + F^{\rm id}(\rho_0$).
The model defined in Eq.~(\ref{eq:pfc1})
looks like a straightforward extension of standard square--gradient Ginzburg--Landau models. It has been formulated
in Ref.~\cite{Eld04}, however, earlier work has established the usefulness of such a free energy
to describe the transition between a homogeneous and a periodically ordered system \cite{Bra75,Swi77}.
For a considerably earlier application to phase transitions in amphiphilic systems, see Ref.~\cite{Gom92}.
Its central features are:
\begin{itemize}
  \item for $c_2,c_4>0$, the term $\propto \phi\nabla^2\phi$ favors a periodically varying $\phi$ and the term
         $\propto \phi\nabla^4\phi$ punishes a spatially varying $\phi$
  \item depending on the parameters, it may have as equilibrium states periodically ordered phases in one dimension (stripes),
        two dimensions (rods) and
        three dimensions (bcc, fcc, hcp)
  \item the characteristic wavenumber of the order parameter field is $q_0 = \sqrt{c_2/(2c_4)}$, which follows from
        $$  \phi(\vect r)\left( c_2\nabla^2 + c_4 \nabla^4  \right)\phi(\vect r) =
           \phi(\vect r)\left( - c_4 q_0^4 + c_4 (q_0^2 + \nabla^2)^2  \right)\phi(\vect r)  $$
\end{itemize}
It turns out that the phase diagram of the above model is equivalent to the formulation of
a reduced model with the free energy according to \cite{Jaa10}
\bea
 \label{eq:fpfc}
  F_{\rm PFC} = \int d^3 x f_{\rm PFC} = \int d^3 x \left( \Psi(\vect x)
                 \left[ -\epsilon + (1+\nabla)^2 \right] \Psi(\vect x) +  \frac{\Psi(\vect x)^4}{4} \right),
\eea
which we call the (actual) phase field crystal (PFC) model. Indeed, we can define the dimensionless coordinate
$\vect x$, a free energy scale $E_0$ and the reduced field $\Psi$ through the transformations
\bea
  \vect x & = & q_0 \vect r = \sqrt{\frac{c_2}{2c_4}} \vect r\;, \nonumber \\
  E_0  &=& \frac{\rho_0^4 c_4^2 q_0^5}{g_{\rm id}} \;, \label{eq:pfctrafo} \\
  \Psi &=& \sqrt{\frac{g_{\rm id}}{\rho_0^2 c_4 q_0^4}}   \left( \Phi - \frac{b_{\rm id}}{3g_{\rm id}} \right)  \; \nonumber
\eea
and the free energy in Eq.~(\ref{eq:pfc1}) becomes
\bea
 \label{eq:pfc2}
  \beta{\cal F}= E_0 \int d^3 x \left( B_0 + B_1 \Psi(\vect x) + f_{\rm PFC}      \right)\;.
\eea
with the value of 
$\epsilon$ (see Eq.~(\ref{eq:fpfc}) for the definition of $f_{\rm PFC}$) given by
\bea
 \label{eq:eps}
   \epsilon = \frac{1}{\rho_0^2 c_4 q_0^4}\left( -a_{\rm id} -\rho_0^2(c_0-c_4 q_0^2) +
    \frac{b_{\rm id}^2}{3g_{\rm id}}  \right) .
\eea
The constants $B_0$ and $B_1$ are given by
\bea
   B_0 &=& \frac{108 g_{\rm id}^3 \beta F_0 + 36 g_{\rm id}^2 b_{\rm id} \rho_0 \beta \mu +
                 b_{\rm id}^2(6a_{\rm id}g_{\rm id} - b_{\rm id}^2 + 6g_{\rm id}\rho_0^2 c_0) }
                {108 g_{\rm id}^2 q_0^8 \rho_0^4 c_4^2}  \;,  \\
   B_1 &=& \sqrt{{\rho_0^2 c_4 q_0^4}g_{\rm id}}\; \frac{27g_{\rm id}^2 \rho_0 \beta \mu +
                  b_{\rm id}(9a_{\rm id}g_{\rm id} -2b_{\rm id}^2+9g_{\rm id}\rho_0^2 c_0)}
                 {27g_{\rm id}^2 q_0^8 \rho_0^4 c_4^2} \;.
\eea
The term $B_0 + B_1 \Psi$ in the free energy Eq.~(\ref{eq:pfc2}) does not influence the location of
the phase boundaries since it is only linear in $\Psi$ but it affects the values of the free energy
density and the chemical potential at coexistence. Since we will determine these values explicitly lateron,
we have given the expressions for $B_0$ and $B_1$ explicitly. Thus the phase diagram
is determined by the variables occuring in $f_{\rm PFC}$, i.e. only by the parameter $\epsilon$ and
$\bar\Psi$, the average value of $\Psi(\vect x)$. The associated phase diagram has been calculated
in Ref.~\cite{Jaa10} and is depicted in Fig.~\ref{fig:phasediag_pfc}.

In the next section, we turn to an exemplification of DFT, Taylor--expanded DFT and PFC for the hard sphere system.
In particular, we will examine minimized solutions for crystals, values for crystal--liquid coexistence
and crystal--liquid surface tensions as well as interfacial profiles for density modes. It will turn out that the
apparent straightforwardness of the PFC derivation from Taylor--expanded DFT is misleading and cannot be upheld,
if one likes to describe crystals with PFC.

\section{Results for hard sphere crystals and crystal--liquid interface}

\label{sec:results}

\subsection{DFT: Fundamental Measure Theory}

\label{sec:fmt_results}

For hard spheres, fundamental measure theory (FMT) allows the construction of very precise functionals
\cite{Ros89,Tar00,Tar08,Rot10}.
Essentially, FMT postulates an excess free energy with a  local free energy
density \tr{expressed} in a set of weighted densities $n_\alpha$:
\bea
 \label{eq:fex_fmt}
  {\cal F}^{\rm ex}[\rho] &=& \int d^3r \Phi(n_\alpha(\vect r)) \;.
\eea
The weighted densities are constructed as convolutions of the density with weight functions,
$n_\alpha(\vect r) = \rho * w^\alpha(\vect r)$.
The weight functions reflect
the geometric properties of the hard spheres.
For one species, the weight functions include four scalar functions
$w^0 \dots w^3$, two vector functions 
$\vect w^1, \vect w^2$ and a tensor function
$w^t$ defined as
\bea
 w^3 = \theta(R-|\vect r|)\;, \qquad
 w^2 = \delta(R-|\vect r|)\;, \qquad w^1 = \frac{w^2} {4\pi R}\;,
  \qquad w^0 = \frac{w^2}{4\pi R^2}\;,
  \nonumber\\
 \vect w^2 =\frac{\vect r}{|\vect r|}\delta(R-|\vect r|)\;,
 \qquad \vect w^1 = \frac{\vect w^2}{4\pi R} \;, \nonumber \\
 w^t_{ij} = \frac{r_i r_j}{\vect r^2} \delta(R-|\vect r|) \;.
\eea
Here, $R=\sigma/2$ is the hard sphere radius. Using these weight functions,
corresponding scalar weighted densities $n_0\dots n_3$, vector weighted densities
$\vect n_1, \vect n_2$ and one tensor weighted density $n_t$ are defined.
In constructing the free energy density $\Phi$, arguments
concerning the correlations in the bulk fluid and arguments for strongly inhomogeneous
systems are used (for reviews see Refs.~\cite{Tar08,Rot10}). For the bulk, $\Phi$ is required to reproduce exactly the second and third
virial coefficent of the direct correlation function. Furthermore,
consistency with a scaled particle argument and/or
imposition of the Carnahan--Starling equation of state leads to
the following form of the excess free energy density \cite{Rot10}:
\bea
 \label{eq:phi_hs}
   \Phi( \{\vect n[\rho (\vect r)]\} ) &=&   -n_0\,\ln(1-n_3) +
      \varphi_1(n_3)\;\frac{n_1 n_2-\vect n_1 \cdot \vect n_2}{1-n_3} +
   \nonumber   \\
      & & \varphi_2(n_3)\; \frac{3 \left( -n_2\, \vect n_2\cdot \vect n_2 + n_{2,i} n_{t,ij} n_{2,j}
        + n_2\,  n_{t,ij}  n_{t,ji} - n_{t,ij}  n_{t,jk} n_{t,ki}
   \right) }{16\pi(1-n_3)^2}\;.
\eea
Here, $\varphi_1(n_3)$ and $\varphi_2(n_3)$ are functions of the local packing density $n_3(\vect r)$.
With the choice
\bea
 \label{eq:frf}
 \varphi_1 = 1\; \text{and } \varphi_2 = 1
\eea
we obtain the Tarazona tensor functional \cite{Tar00} which is built upon the original
Rosenfeld functional \cite{Ros89}. The latter gives the fluid equation of state and
pair structure of the Percus--Yevick approximation. Upon setting
\bea
 \label{eq:fwbII}
 \varphi_1 & = & 1 + \frac{2n_3-n_3^2 + 2(1-n_3) \ln(1-n_3 )}{3n_3} \\
 \varphi_2 & = & 1 - \frac{2n_3-3n_3^2 + 2n_3^3 + 2(1-n_3 )^2 \ln(1-n_3 ) }
                          {3 n_3^2}  \nonumber
\eea
we obtain the tensor version of the recently introduced White Bear II (WBII) functional \cite{Han06}.

For bulk crystals, very accurate free energy results can already be obtained using a Gaussian approximation
for the density,
\bea
  \label{eq:Gauss_ansatz}
  \rho(\vect r) &= & \sum_{{\rm lattice\;sites}\;i} (1-n_{\rm vac})\,\left(\frac{\alpha}{\pi}\right) ^{\frac{3}{2}}
  \,\exp\left( -\alpha (\vect r - \vect r_i)^2/\sigma^2
   \right) \;,
\eea
and minimizing the total free energy with respect to the width parameter $\alpha$ and the vacancy concentration
$n_{\rm vac}$ at a fixed bulk density. At coexistence $\alpha \sim 80$, in accordance with simulations
 and free energies per particle between Gaussian parametrized DFT and simulations agree on the level of 0.01 $\kt$
\cite{Oet10}.
However, the Tarazona functional (\ref{eq:frf}) yields $n_{\rm vac} \to 0$ while the WBII functional
Eq.~(\ref{eq:fwbII}) gives a finite
equilibrium vacancy concentrations  $n_{\rm vac} = O(10^{-5})$ which is about a factor 10 smaller
than in the simulation results \cite{Kwa08}. This fine difference has important consequences:
performing an unconstrained minimization (see Eq.~(\ref{eq:minimizingF}) with vanishing external potential)
only the WBII functional gives an absolute minimum for the free energy with a value for the chemical
potential which is consistent with the \tr{derivative of the crystal free energy density with respect to the
bulk density}
(see Ref.~\cite{Oet10} for further details). This implies that a free minimization for the crystal--fluid
interface can only be performed using the WBII functional.

The free minimization of the crystal--fluid interface is a non--trivial task, 
a brief description
of the method (also applicable in the case of Taylor--expanded DFT) is given in App. \ref{app:dftmin}.
Results for the surface tension with crystal faces in different \tm{orientations}
have been reported in Ref.~\cite{Har12} and are also given in Table~\ref{tab:results}.
There is agreement with simulations in the ordering $\gamma_{[100]}>\gamma_{[110]}>\gamma_{[111]}$ and as far as the accuracy
of the data permits, also in the values of  the relative anisotropies (i.e. the values of
$(\gamma_{[100]}-\gamma_{[110]})/\gamma_{[100]}$ and $(\gamma_{[100]}-\gamma_{[111]})/\gamma_{[100]}$).
There is no clear consensus between different simulation methods on the absolute values of the $\gamma$'s,
the latest results are closer to the DFT values, however.

Overall these results corroborate what is known from many applications of FMT on (dense) liquids
\cite{Rot10,Oet09,Bot09}: it is a quantitative theory and the accuracy also extends to the description
of crystalline systems.
Therefore we can consider FMT as a benchmark theory against which subsequent approximative approaches should be tested.

\begin{table}

\begin{tabular}{l|l|ll|lll|l}

\hline \hline

                   & FMT    & T--DFT & T--DFT &  & PFC & & SIM  \\
                   & (WBII) & (PY)        & (WBII)      & & &      &   \\
$\rho_0\sigma^3$           &        & 0.9461      & 1.026       & & 0.94 &     &   \\
$\epsilon$         &        &             &             & 0.50 & 0.65 & 0.80 & \\ \hline
$\rho_{\rm cr}\sigma^3$    & 1.039  & 1.049       & 1.123       &      & {\em 1.04} &   & 1.041 \cite{Zyk10} 1.039 \cite{Dav10} \\
$\rho_{\rm fl}\sigma^3$    & 0.945  & 0.944       & 1.021       &      & {\em 0.94} &   & 0.940 \cite{Zyk10} 0.938 \cite{Dav10}  \\
$(\beta F/N)_{\rm cr}$   & 4.96   & 5.33        & 7.23        &      & 5.20 &         & 4.96$^\dag$   \\
$(\beta F/N)_{\rm fl}$   & 3.82   & 3.99        & 5.05        &      & 3.93 &         & 3.75$^\S$ \\
$\beta\mu_{\rm coex}$   & 16.38  & 17.44       & 21.51       &      & 17.16 &        & 16.09$^\S$ \\
$n_{\rm vac}$      & $2\cdot 10^{-5}$ & 0.10 & 0.09 & $-0.11$ & $-0.12$ & $-0.13$ & $3\cdot 10^{-4}$ \cite{Kwa08}  \\ \hline
$\beta\sigma^2\gamma_{[100]}$     & 0.69 \cite{Har12}   & 0.89        & 1.31  & 0.140 & 0.074 & 0.046 & 0.58 \cite{Dav10} 0.63 \cite{Har12} 0.64 \cite{Fer11} \\
$\beta\sigma^2\gamma_{[110]}$     & 0.67                & 0.85        & 1.21  & 0.132 & 0.070 & 0.043 & 0.56 \phantom{[20]} 0.61 \\
$\beta\sigma^2\gamma_{[111]}$     & 0.64                & 0.78        & 1.09  & 0.105 & 0.055 & 0.034 & 0.54 \phantom{[20]} 0.60 \\
\hline \hline

\end{tabular}
\begin{flushleft}
$^\dag$ Free energy for $\rho_{\rm cr}\sigma^3=1.041$ \cite{Zyk10} using an improved fit in the form of the Speedy
equation of state from Ref.~\cite{Oet10}. \\
$^\S$ Free energy and chemical potential for $\rho_{\rm fl}\sigma^3=0.940$ \cite{Zyk10} from the Carnahan--Starling equation of state.
\end{flushleft}

\caption[]{Results for coexistence properties and crystal--fluid surface tension using the different approaches
considered in this work. The PFC results have been obtained using the fitting procedure described in
Sec.~\ref{subsec:pfc_results} for the reference density $\rho_0\sigma^3=0.94$ with the crystal and fluid coexistence densities
as input (in italics).}
\label{tab:results}

\end{table}

\subsection{Taylor--expanded DFT}

\label{sec:taylor_results}

We consider the Taylor--expanded functional Eq.~(\ref{eq:f_hnc}) of the FMT functionals, Eq.~(\ref{eq:frf}) (Tarazona functional)
and Eq.~(\ref{eq:fwbII}) (WBII functional).
The nontrivial, second--order term in the functional involves the
direct correlation function $c^{(2)}(r;\rho_0)$. It is the second derivative of the excess
free energy functional and is given in both cases by the polynomial form
\bea
 \label{eq:c2power}
  c^{(2)}(r;\rho_0) &=& (a_1 + a_2 r/\sigma + a_3 (r/\sigma)^3) \theta(\sigma-r) \;.
\eea
Using the packing fraction $\eta=\pi\sigma^3/6\,\rho_0$, the coefficients for $c^{(2)}$ of the Tarazona
functional are those of the famous Percus--Yevick solution,
\bea
  a_1 &=&  - \frac{(1 + 2 \eta)^2}{(1 - \eta)^4}\;, \nonumber \\
  a_2 & = &   6 \eta \frac{(1 + \eta/2)^2}{(1 - \eta)^4}\;, \qquad \mbox{(Tarazona)} \label{eq:c2tar}\\
  a_3 & = &  \frac{\eta}{2}  a_1 \;. \nonumber
\eea
The coefficients for  $c^{(2)}$ of the WBII functional are given by
\bea
  a_1 &=& -\frac{1 + 4\eta + 4\eta^2 - 4\eta^3 + \eta^4}{(1-\eta)^4}\;, \nonumber  \\
  a_2 & = & \frac{-2+25\eta+12\eta^2-10\eta^3+2\eta^4}{3(1-\eta)^4} +
            \frac{2\ln(1-\eta)}{3\eta}\;  \qquad \mbox{(White Bear II)} \label{eq:c2wbII}\\
  a_3 & = & \frac{1-4\eta+2\eta^2-3\eta^3+\eta^4}{(1-\eta)^4}+\frac{\ln(1-\eta)}{\eta}  \;. \nonumber
\eea

\begin{figure}
 \epsfig{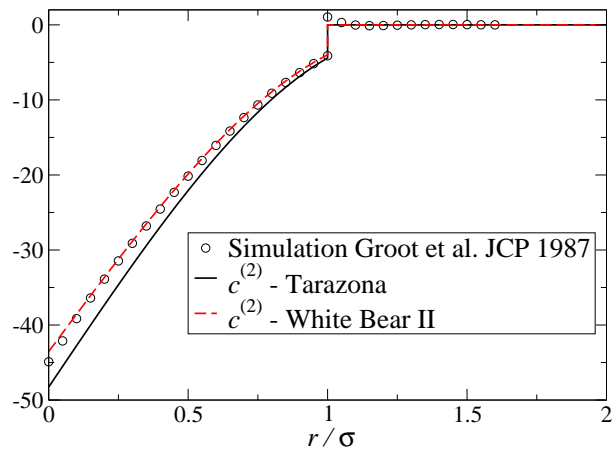}
 \caption{(color online) Direct correlation function for hard spheres at a density $\rho\sigma^3=0.9$, somewhat below
   freezing. Simulation data are taken from Ref.~\cite{Gro87}.
 }
 \label{fig:c2}
\end{figure}

Both forms approximate the simulation results for $c^{(2)}$ reasonably well, 
see Fig.~\ref{fig:c2},
with the WBII form matching \tb{ more closely}.
As we will see, this does {\em not} imply better
results for the Taylor--expanded functional in the WBII case.

The Taylor--expanded functional contains the reference density $\rho_0$  as an additional parameter.
It would be desirable to choose it such that it also recovers the fluid density at coexistence. In this way it is guaranteed that at least the fluid properties are almost exact
if the Taylor--expanded functional is fixed adequately.
In order to determine $\rho_0$ with moderate effort,
we employ the approach already taken in Ref.~\cite{Jon85}. For a given
bulk density $\rho_{\rm b}$, we parameterize the density profile in the Gaussian form
Eq.~(\ref{eq:Gauss_ansatz}) and determine the three free parameters $\alpha$ (Gaussian width),
$n_{\rm vac}$ (vacancy concentration) and $\rho_0$ by minimizing the
functional for the grand potential {\em difference} per particle between crystal and liquid at the
chemical potential $\mu(\rho_0)$:
\bea
  \frac{\beta \Delta \Omega(\rho_{\rm b})}{N} &= & \mbox{min}_{\alpha, n_{\rm vac},\rho_0}
   \left[ \frac{1}{\rho_{\rm b} V}
    \int_V d^3 r \left( \rho(\vect r) \ln\frac{\rho(\vect r)}{\rho_0} - \rho(\vect r) + \rho_0
\right) - \right. \nonumber \\
  & & \qquad\qquad\left. - \frac{1}{2\rho_{\rm b} V} \int_V d^3 r d^3 r'
    c^{(2)}(\vect r, \vect r';\rho_0) \Delta\rho(\vect r)\Delta\rho(\vect r')\right] \;,
\eea
where $\Delta\rho(\vect r) = \rho(\vect r) - \rho_0$ is defined as before. In practice, the space integrations
can be restricted to one cubic fcc unit cell. Finally, the bulk density $\rho_{\rm b}$ is varied
until $\Delta \Omega(\rho_{\rm b})=0$, i.e. at this point $\rho_{\rm b}$ and $\rho_0$ (from the
minimization) correspond to the coexisting crystal and fluid densities within the Gaussian
approximation. We pick this $\rho_0$ as the reference density for the unconstrained minimization, and
re--determine the bulk crystal and fluid densities at coexistence through a full minimization
and a subsequent Maxwell construction. These densities are only slightly shifted from the
ones obtained within the Gaussian approximation (see Table~\ref{tab:results}).

The results are partially surprising (see Table~\ref{tab:results}). For the Taylor--expanded
Tarazona functional (i.e. employing the $c^{(2)}$ of Percus--Yevick), the coexisting densities
are still very close to the simulation and FMT (0.944 and 1.049). In the Gaussian approximation,
similar values have been already obtained in 1985 by Jones and Mohanty \cite{Jon85}. The
crystal free energy is too big ($\beta F/N=5.33$ vs. 4.96 from simulation and FMT) and the
width of the Gaussian peaks is much too narrow ($\alpha\sim 600$ vs. 80 from simulation and FMT).
For the WBII functional (i.e. the better functional, with a more precise $c^{(2)}$) the coexistence
densities are considerably off (1.021 and 1.123), consequently the crystal free energy is too big by
40\% and the Gaussian width parameter $\alpha \sim 1000$ stands for even narrower peaks.
For both functionals, the vacancy concentrations are too large by 3 orders of magnitude.

As before for FMT, we determine the surface tensions using a full minimization
(see App.~\ref{app:dftmin}). It is gratifying that the
ordering $\gamma_{[100]}>\gamma_{[110]}>\gamma_{[111]}$ is upheld but the relative anisotropies
are too large by approximately a factor of 2 and the average value of the
surface tension is too large (Taylor--expanded Tarazona functional: 0.84,
Taylor--expanded White Bear II functonal: 1.2 vs. 0.67 from full FMT, all values in units of $1/(\beta\sigma^2)$).

In conclusion, it is apparent that through Taylor expansion of the precise FMT functionals,
crystal free energies, surface tensions and vacancy concentrations are severely affected.
Nevertheless, a qualitative descriptions is still achieved. The reason for the quantitative
failure of the Taylor--expanded functionals is most likely due to the fact that the packing
constraints or free energies for highly localized states are not respected very well.
This is in contrast to the FMT functionals which have incorporated the correct description of  localized states
\cite{Tar00,Tar08}.

\subsection{Phase-field crystal modelling}
\label{subsec:pfc_results}

\subsubsection{Fixing parameters using bulk properties}

From the phase diagram of the PFC model (Fig.~(\ref{fig:phasediag_pfc})) with free energy density
$$f_{\rm PFC} =\Psi(\vect x) \left[ -\epsilon + (1+\nabla)^2 \right] \Psi(\vect x) +  \frac{\Psi(\vect x)^4}{4}$$
one infers that, for the description
of stable fluid--crystal(fcc) coexistence, a parameter range of $\epsilon \sim 0.5...1$ is necessary.
For lower $\epsilon$, fcc is only metastable with respect to bcc, and for $\epsilon <0$ there are no ordered
phases at all. However, following the derivation of PFC from the Taylor--expanded functional, one is led
to the free energy in Eq.~(\ref{eq:pfc1}) with coefficients $a_{\rm id}$, $b_{\rm id}$, $g_{\rm id}$ from the
expansion of the ideal gas free energy and the Taylor--coefficients $c_0$, $c_2$, $c_4$ from the expansion
of the Fourier transform of the direct correlation function $\tilde c^{(2)} (k;\rho)$. Using the explicit form
for $c^{(2)}(r) = a_1 + a_2 (r/\sigma) + a_3 (r/\sigma)^3$ $(r/\sigma<1)$, we find
\bea
  c_0 &=& 4\pi\left(\frac{a_1}{3} + \frac{a_2}{4} + \frac{a_3}{6}\right)\;, \nonumber \\
  c_2 &=& \frac{4\pi}{6}\left(\frac{a_1}{5} + \frac{a_2}{6} + \frac{a_3}{8}\right)\;, \\
  c_4 &=& \frac{4\pi}{120}\left(\frac{a_1}{7} + \frac{a_2}{8} + \frac{a_3}{10}\right)\;, \nonumber
\eea
and inserting into the equation (\ref{eq:eps}) for $\epsilon$ we obtain $\epsilon=-0.6 \dots -0.7$
($\rho_0\sigma^3=0.94 \dots 1.04$, Percus--Yevick direct correlation function). Thus, the naive gradient expansion of the
Taylor--expanded functional produces a free energy which shows no sign of a liquid--solid transition!
The reason is essentially that the gradient expansion roughly approximates the Fourier
transform of the direct correlation function $\tilde c^{(2)} (k;\rho)$ and consequently also  the structure
factor defined by
\bea
 \label{eq:sf}
   S(k;\rho) =  \frac{1}{1-\rho\tilde c^{(2)} (k;\rho)} \;.
\eea
The gradient expansion leads to a structure factor which
clearly violates the empirical Verlet--Hansen freezing criterion (height of first peak in $S(k) \agt 2.8$)
whereas $S(k)$ from the PY direct correlation function fulfills it.

Such a failure of the naive gradient expansion has been noted and discussed before in a case study on
the applicability of PFC for bcc metals \cite{Jaa09}. The remedy proposed was to fit $c_0$, $c_2$, $c_4$
to the first maximum of $\tilde c^{(2)}(k;\rho)$ (or $S(k;\rho)$) at around $k\sigma \sim 7$ and also to fit the coefficients
$a_{\rm id}$, $b_{\rm id}$, $g_{\rm id}$ in order to achieve a reasonable description of coexistence.
The results from this procedure can be considered as partially successful: the correct description
of the first peak of the structure factor needs a value for $c_0$ which is  too small and hence causes deviations of
the liquid isothermal compressibility $\beta (\partial p/\partial\rho)^{-1}=1/(1-\rho \tilde c^{(2)} (0;\rho)) = S(0;\rho)$. Also, bulk free energies are not well captured \cite{Jaa09}.

\begin{figure}
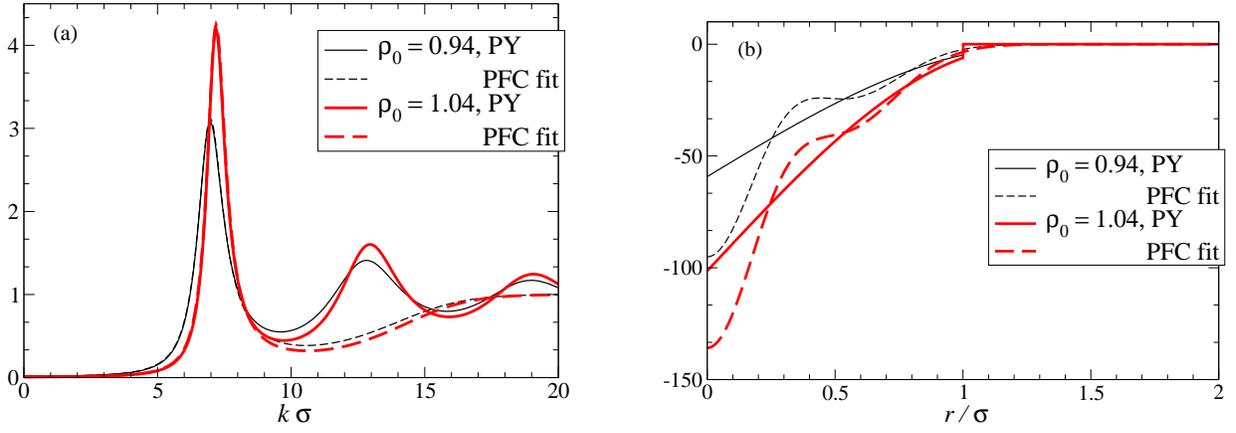

  \epsfig{file=fig3a.eps, width=7.5cm} \hspace{1cm}
  \epsfig{file=fig3b.eps, width=7.5cm}
  \caption{(color online) Comparison between Percus--Yevick results and PFC fits. (a) Structure factor $S(k;\rho_0)$. (b)
  Direct correlation function $c^{(2)}(r;\rho_0)$.}
  \label{fig:cfit}
\end{figure}

Another serious problem is related to the identification of the order parameter $\Phi(\vect r)=\rho(\vect r)/\rho_0-1$
(see Eq.~(\ref{eq:pfc1})) with the shifted  dimensionless density. The order parameter $\Psi$  of  the PFC model is related to a rescaling and shift of the order parameter $\phi$.
Numerical solutions of the PFC model for $\epsilon >0$ show that the order parameter solutions for bulk crystals
can still be approximated by Gaussians (Eq.~(\ref{eq:Gauss_ansatz})), but they are much more spread out (width parameter $\alpha \sim 10$,
compared to 80 (for the case of FMT) and 500\dots 1000 (for the case of Taylor--expanded DFT). Consequently $\Phi(\vect r)$ has to be interpreted rather as a
smeared--out reduced density. We will model this idea using a simple, normalized  Gaussian smearing function
with width $\alpha'$ leading to
the following reinterpretation of $\Phi(\vect r)$:
\bea
  \Phi(\vect r) & =  &\frac{\bar\rho(\vect r)}{\rho_0} -1 \;, \\
  \label{eq:rhobar}
   \text{with }\bar\rho(\vect r) &= & \int d\vect r' \left(\frac{\alpha'}{\pi}\right)^{\frac{3}{2}} \exp(-\alpha' r'^2/\sigma^2) \rho(\vect r-\vect r')\;.
\eea
Inserting this {\em ansatz} into the approximative free energy in Eq.~(\ref{eq:pfc1}) and applying $c^{(2)}=-\delta {\cal F}^{\rm ex}/(\delta\rho\delta\rho)$, we find
the following Fourier
transform for the direct correlation function:
\bea
  \label{eq:c2k_pc}
  \tilde c^{(2)}(k;\rho) &=& \exp\left(-\frac{k^2}{2\alpha'} \right)(-c_0 + c_2 k^2 -c_4 k^4 \dots) \;.
\eea
Treating $\alpha',c_0,c_2,c_4$ as fitting parameters, we obtain very accurate fits of {\em both}, $\tilde c^{(2)}(k;\rho)$ and $S(k)$
in the ``long wavelength'' region $k\sigma \alt 10$
for different choices of the reference density $\rho_0$, see Fig.~\ref{fig:cfit} and Table~\ref{tab:pfcpars}.
The good matching properties are a result of the fit \tb{ with} $\alpha'\approx 14$ This value naturally accounts for the
extended width of the bulk crystal solutions for $\Phi(\vect r)$.
Additionally we also have the qualitatively correct behavior for $c^{(2)}(r)$ for {\em short} distances $r \alt \sigma$
(see Fig.~\ref{fig:cfit}).
 Furthermore, in line with the previous studies
\cite{Wu07,Jaa09}, we treat the coefficients $a_{\rm id}, b_{\rm id}, g_{\rm id} \to a,b,g$ as free parameters since the ideal gas free energy can not be applied to a smoothed density field $\bar\rho$ as defined above.
A direct way to fix these parameters is by using the physical hard sphere coexistence densities
$\rho_{\rm cr}$ and $\rho_{\rm fl}$ as input.
Since the PFC phase diagram is described
by the reduced PFC free energy in Eq.~(\ref{eq:fpfc}), see Fig.~\ref{fig:phasediag_pfc}, the triple
$\epsilon, \bar\Psi_{\rm cr}(\epsilon),\bar\Psi_{\rm fl}(\epsilon)$ fixes the three coefficients $a,b,g$
using the third  relations of Eq.~(\ref{eq:pfctrafo}) and Eq.~(\ref{eq:eps}).
Here, $\bar\Psi_{\rm cr}(\epsilon)$ and $\bar\Psi_{\rm fl}(\epsilon)$ are the coexistence values for the average order parameter
of the fcc crystal and of the fluid, respectively.

In summary, a reasonable procedure to fix the PFC parameters is as follows:
\begin{itemize}
 \item fix a reference density $\rho_0$, fit $S(k;\rho_0)=1/(1-\rho_0\tilde c^{(2)}(k;\rho_0))$ using Eq.~(\ref{eq:c2k_pc})
   in the wave vector region including the first peak of the structure factor ($k\sigma \alt 10$).
 \item fix the PFC parameter $\epsilon$ and require coexistence at the physical coexistence densities: this determines
     $a,b,g$ and consequently also the length scale $q_0$ and the free energy scale $E_0$ of the PFC model
\end{itemize}
We have gathered the results of this procedure for some combinations of reference densities and $\epsilon$--parameters in
Table~\ref{tab:pfcpars}. Note that the fitted values of $a,b,g$ are one to two orders of magnitude larger than the
ideal gas values $a_{\rm id}=\rho_0$, $b_{\rm id}=\rho_0/2$, $g_{\rm id}=\rho_0/3$. This is a consequence of the
fact that the PFC order parameter should be considered as a smeared--out density.

\begin{table}

\begin{tabular}{ll|llll|llll|llll}
\hline \hline
$\rho_0\sigma^3$  & $\epsilon$ &   $\alpha'$    & $c_0\sigma^3$ & $c_2\sigma^5$ & $c_4\sigma^7$ & $a\sigma^3$ & $b\sigma^3$ & $g\sigma^3$ &  $E_0$  & $\left(\frac{\beta F}{N}\right)_{\rm cr}$ & $\left(\frac{\beta F}{N}\right)_{\rm fl}$ & $\beta \mu_{\rm coex}$ & $n_{\rm vac}$   \\[2mm]  \hline
\multicolumn{2}{c}{{\em Simulation}} &   &       &       &        &       &       &       &      & 3.75 & 4.96 &  16.1  & 10$^{-4}$ \\ \hline
0.94     & 0.5      &    14.21          & 61.93 & 2.567 & 0.0249 & 23.45 & 69.75 & 33.13 & 0.279 & 3.93 & 5.20 & 17.16 & -0.11  \\
         & 0.65     &                   &        &      &        & 32.40 & 112.0 & 62.75 & 0.147 & 3.93 & 5.20 & 17.16 & -0.12 \\
         & 0.8      &                   &        &      &        & 42.02 & 160.5 & 101.0 & 0.091 & 3.93 & 5.20 & 17.16 & -0.13 \\ \hline
1.0      & 0.5      &    13.93         &  80.36 & 3.201 & 0.03013 & 21.04 & 98.19 & 54.57 & 0.342 & 3.99 & 5.06 & 15.06 & -0.06 \\
         & 0.65     &                  &        &       &         & 26.75 & 154.9 & 103.4 & 0.181 & 4.01 & 5.01 & 14.48 & -0.07 \\
         & 0.8      &                  &        &       &         & 32.42 & 218.6 & 166.3 & 0.112 & 4.02 & 4.97 & 13.90 & -0.08 \\ \hline
1.04     & 0.5      &    13.99         &  96.56 & 3.745 & 0.03458 & 16.58 & 122.9 & 76.15 & 0.397 & 4.11 & 4.94 & 12.77 &-0.03 \\
        & 0.65      &                  &        &       &         & 18.40 & 191.1 & 144.2 & 0.209 & 4.15 & 4.89 & 11.92 & -0.04 \\
        & 0.8       &                  &        &       &         & 19.53 & 266.5 & 232.1 & 0.130 & 4.18 & 4.84 & 11.07 & -0.05 \\
\hline \hline
\end{tabular}

\caption{PFC fitting parameters and results for the bulk liquid and crystal phases at coexistence.}
\label{tab:pfcpars}

\end{table}

In Table \ref{tab:pfcpars} we also give the free energy per particle for the coexisting bulk liquid and crystal phases, their
chemical potential and the vacancy concentration of the coexisting crystal, obtained from a minimization of the PFC free energy of Eq.~(\ref{eq:fpfc}) for bulk crystal states. \tm{.}
For the absolute values of the free energy and the coexisting chemical potential, one needs to determine the constant
and linear terms in the PFC order parameter $\Psi$ (see Eq.~(\ref{eq:pfc2})\tm{)}. We find that, for the reference density
$\rho_0 \sigma^3=0.94$, the coexistence free energies and the chemical potential are well recovered
in this case $\rho_0 = \rho_{\rm fl}$ , the liquid free energy and the chemical potential are those of the
PY theory and are hence reasona\tm{b}ly accurate. The good approximation of the crystal--fluid free energy difference
is  in contrast to the findings in Ref.~\cite{Jaa09}.

\begin{figure}
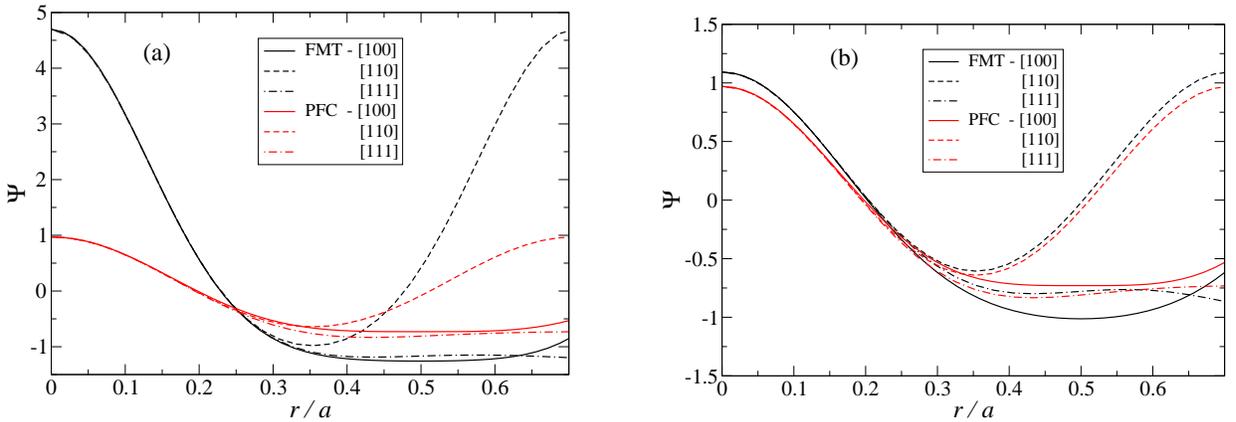

  \epsfig{file=fig4a.eps, width=7.5cm} \hspace{1cm}
  \epsfig{file=fig4b.eps, width=7.5cm}
  \caption[]{Comparison between FMT and PFC for the order parameter $\Psi$ in the bulk crystal at coexistence,
  for three different lattice directions in the fcc cubic unit cell of side length $a$. The order parameter
  is calculated from the FMT density profile $\rho(\vect r)$ by $\Psi(\vect r) = \sqrt{g/(\rho_0^2 c_4 q_0^2)} (\bar\rho(\vect r)/\rho_0-1-
  b/(3g))$ with $q_0^2=c_2/(2c_4)$ and $\bar\rho(\vect r)$ given by the convolution in Eq.~(\ref{eq:rhobar}).
  Values for $b,g,c_2,c_4$  are optained for the data set $\rho_0\sigma^3=0.94$ and $\epsilon=0.5$ (see Table~\ref{tab:pfcpars}), and
  the smearing width for $\bar\rho$ is $\alpha'=14.0$ in (a) and $\alpha'=7.0$ in (b). }
  \label{fig:psi_comparison}
\end{figure}

However, the order parameter profiles $\Psi$ from FMT and PFC in the crystal unit cell at coexistence agree
only qualitatively, see Fig.~\ref{fig:psi_comparison}. Here, the FMT solution for $\Psi$ is calculated via
the third expression of Eq.~(\ref{eq:pfctrafo}), but with the smeared density profile defined in Eq.~(\ref{eq:rhobar}).
Using a smearing parameter $\alpha'=14.0$, consistent with the structure factor fit in Table~\ref{tab:pfcpars},
yields order parameters that vary \tb{ more strongly} in FMT than in PFC (Fig.~\ref{fig:psi_comparison}(a)).
Although in Fig.~\ref{fig:psi_comparison}(a) the comparison is shown only for one choice
of $\rho_0$ and $\epsilon$, the differences between FMT and PFC are also found for other parameter combinations.
Only by choosing $\alpha'=7.0$ (stronger smearing), the order parameter profiles agree almost quantitatively
(see Fig.~\ref{fig:psi_comparison}(b)). For such low values of $\alpha'$, the  fitting procedure  gives too large values for
the inverse PFC length scale $q_0$ and too small values for the free energy scale $E_0$. Thus, the
order parameter description in PFC is somewhat defective.

The relative vacancy concentration $n_{\rm vac}$ can be calculated via the number of
particles $N_{\rm cell}=\rho_{\rm cr} (a_{\rm min}/q_0)^3$ in the fcc cubic unit cell by
$n_{\rm vac}= 1-N_{\rm cell}/4$. Here, $a_{\rm min}$ is the cubic unit cell length in dimensionless
PFC units which follows from minimizing the PFC free energy density at the value for the average
order parameter $\bar\Psi$ at coexistence. The resulting $n_{\rm vac}$ is negative (see Table~\ref{tab:pfcpars})
and of order $0.1$ which  implies a considerable concentration of interstitial particles.
This is, of course, unphysical but it simply follows from fixing PFC coexistence to
the correct physical densities. This observation reflects on\tm{c}e more the difficulties in fixing parameters.
Similarly to the case of the Taylor--expanded functional, it can not be expected that a ``generic''
free energy functional like the PFC functional can capture the correlations in the nearest--neighbor
shell of a crystalline particle correctly, especially the condition of no overlap between particles.
These correlations determine the precise value of $n_{\rm vac}$.

\subsubsection{Crystal--fluid surface tensions}

We have determined the equilibrium order parameter profile and the associated PFC free energy
for the crystal--fluid interface as the long--time limit of the solution to the  dynamic equation
\bea
 \label{eq:pfc_dyn_eq}
  \frac{\partial\Psi(\vect x, t)}{\partial t} = \nabla^2 \frac{\delta F_{\rm PFC}}{\delta\Psi(\vect x,t)}
\eea
(see Eqs.~(\ref{eq:ddft1}) and (\ref{eq:ddft2})), with initial conditions given by a trial profile
for a crystal slab in the simulation box filled otherwise with liquid. Any mobility coefficient, relating the
PFC time $t$ in the above equation to real time, is unimportant for the discussion of equilibrium properties.
The order parameter profiles and the PFC free energy can be converted to density profiles and physical free
energies using Tab.~\ref{tab:pfcpars}.
 We have noted a certain sensitivity
of the surface tensions and the order parameter profiles to the grid spacing and the precise extensions of the
simulation box. These details are discussed in App.~\ref{app:pfcmin_interface}.

\begin{figure}
 \centerline{\epsfig{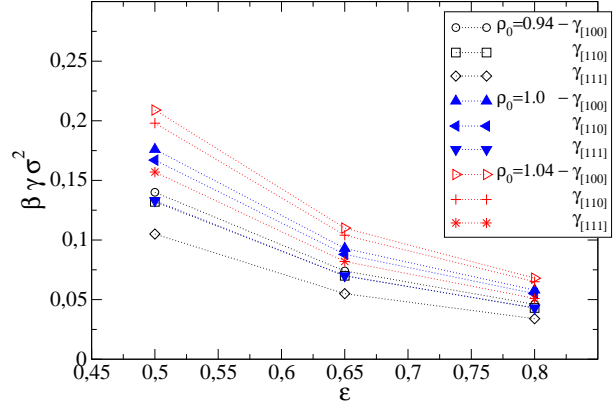}}
 \caption{(color online) Surface tension from the PFC model for different orientations and different reference
   densities $\rho_0$. For different $\rho_0$, the physical surface tensions $\gamma=\hat\gamma q_0^2 E_0$ differ
   since the free energy scale $E_0$ and the inverse length scale squared $q_0^2=c_2/(2c_4)$ differ
  (see Tab.~\ref{tab:pfcpars}). The dimensionless PFC surface tension $\hat\gamma$ is only a function of $\epsilon$
   (values given in text).  }
 \label{fig:gamma_pfc}
\end{figure}

We have calculated the dimensionless PFC surface tension $\hat\gamma$ for the three different orientations
[100],[110] and [111], each for the values of the PFC parameter $\epsilon=0.5$, 0.65 and 0.8. In this
sequence of orientations the results are: 0.0097, 0.0092, 0.0073 ($\epsilon=0.5$), 0.0132, 0.0129, 0.0102 ($\epsilon=0.65$)
and 0.0165, 0.0163, 0.0129 ($\epsilon=0.8$). The conversion to physical surface tensions of the hard sphere
system is given by  $\gamma=\hat\gamma q_0^2 E_0$ with the free energy scale $E_0$ and the inverse length scale squared $q_0^2=c_2/(2c_4)$
given in Tab.~\ref{tab:pfcpars}. This results in different $\gamma$ for different reference densities $\rho_0$ which are
depicted in Fig.~\ref{fig:gamma_pfc}. For the reference density $\rho_0\sigma^3=0.94$ (where PFC bulk crystal data are in good
agreement with FMT and simulation) the surface tension values are given in Tab.~\ref{tab:results} as well.
The physical surfac\tm{e} tensions are largest for highest reference density $\rho_0$ and the lowest PFC parameter $\epsilon$ and
decrease for decreasing $\rho_0$ and/or increasing $\epsilon$. Still, on average for the three different orientations
the surface tensions are too low compared to FMT/simulation values between a factor of about  3 ($\rho_0\sigma^3=1.04$, $\epsilon=0.5$)
and about 15  ($\rho_0\sigma^3=0.94$, $\epsilon=0.8$). The ordering of surface tensions $\gamma_{[100]}>\gamma_{[110]}> \gamma_{[111]}$
is correct for the PFC results but $\gamma_{[111]}$ is smaller than $\gamma_{[100]}$ by about 30\% which differs
considerably from the 5...8\% as found in simulation and FMT. Likewise a strong qualitative difference between the order parameter profiles
of the [111] interface compared to the [100], [110] interfaces is also found: the width of the [111] interface is considerably wider
than of the [100], [110] interfaces (see below). This feature is not present in FMT.





\subsection{Density and order parameter modes at the crystal--fluid interface}

\subsubsection{General theory and previous FMT results}

\label{subsec:mode_theory}

Consider a generic field $\psi(x,y,z)$ which describes the crystal--fluid interface with interface normal
in $z$--direction. In DFT, this field  \tm{is} the
density  $\rho(x,y,z)$ and in PFC, it is the order parameter field $\Psi(x,y,z)$. We can parametrize
the field $\psi$ in terms of a modified Fourier expansion
\bea
 \psi(x,y,z) = \sum_j \exp(\imag \vect K_j \cdot \vect r)\; p_j(z)\;,
\eea
where $\vect K_j$ denotes the reciprocal lattice vector (RLV), $j$ and
the $z$--dependent Fourier amplitude $p_j(z)$ are modes of the field.
One expects that upon
crossing the interface  from the crystal side, all $p_j(z)$ relax to zero for nonzero $\vect K_j$. Only
for $\vect K_j \equiv 0$, the value for the associated mode crosses from the average field $\bar\psi$ of the crystal to
the average field of the fluid.
It is convenient to group the $\vect K_j$ in shells with index $m$,
where all $\vect K_j$ belonging to one shell, can be transformed into each other using the discrete symmetry group
of the crystal under consideration. Thus in the bulk crystal, all $p_j(z) \equiv P_j$
associated with these $\vect K_j$ are equal.
For the fcc crystal, the reciprocal lattice is of bcc
symmetry. We assume $a$ to be the side--length of the cubic unit cell of {\em fcc} and correspondingly $b=2\pi/a$ the
side length of the cubic unit cell of $bcc$ in reciprocal space. The reciprocal basis is given
in Cartesian coordinates, where the axes span the cubic unit cell in reciprocal space, by
$\vect B_1 = b(1,1,-1)$, $\vect B_2 = b(1,-1,1)$ and $\vect B_3 = b(-1,1,1)$. An arbitrary RLV is a linear combination
of the $\vect B_i$. The shells are characterized by a triple $(m,n,k)$ of natural numbers and the $K_j$
belonging to this shell have Cartesian components $b(\pm m, \pm n, \pm k)$ and permutations thereof. Thus, if
$m,n,k$ are mutually distinct, there is a maximum of 48 RLV in one shell. The shells with lowest modulus are given
by $(1,1,1)$, $(2,0,0)$ and $(2,2,0)$.
A listing of the RLV triples up to shell 15 is given in Ref.~\cite{Hay83} (Table I).
At the interface, the
degeneracy of the RLV in one shell is lifted, and we introduce an index $n$ which
distinguishes the possible values of the $z$--component of the RLV. Thus the decomposition becomes
\bea
 \label{eq:int_deco}
 \psi(x,y,z) = \sum_{mn} \sum_j p_{mn}(z)  \exp\left(\imag (\vect K_j)_{mn} \cdot \vect r\right)\;.
\eea
The sum over $j$ is  only for those RLV within shell $m$ which have a common value of $z$--component, as expressed
by the index $n$. In the literature, such a decomposition has been used to parameterize the full 3d density profile
using  only the leading mode in order to facilitate a simplified
order parameter description of the crystal--fluid interface. In the context of PFC, the leading--mode picture has been advanced by Karma {\em et al.} \cite{Wu06}.
If the $z$--component of $(\vect K_j)_{mn}$ is zero,
the mode will be purely real and if that $z$--component is nonzero, the mode will be in general complex
and we denote by $p_{mn}^+(z)$ its real part and by $p_{mn}^-(z)$ its imaginary part.
The $p_{mn}^- (z)$ have the obvious interpretation of {\em phase shifts} of the associated field
oscillations across the interface.

\begin{figure}
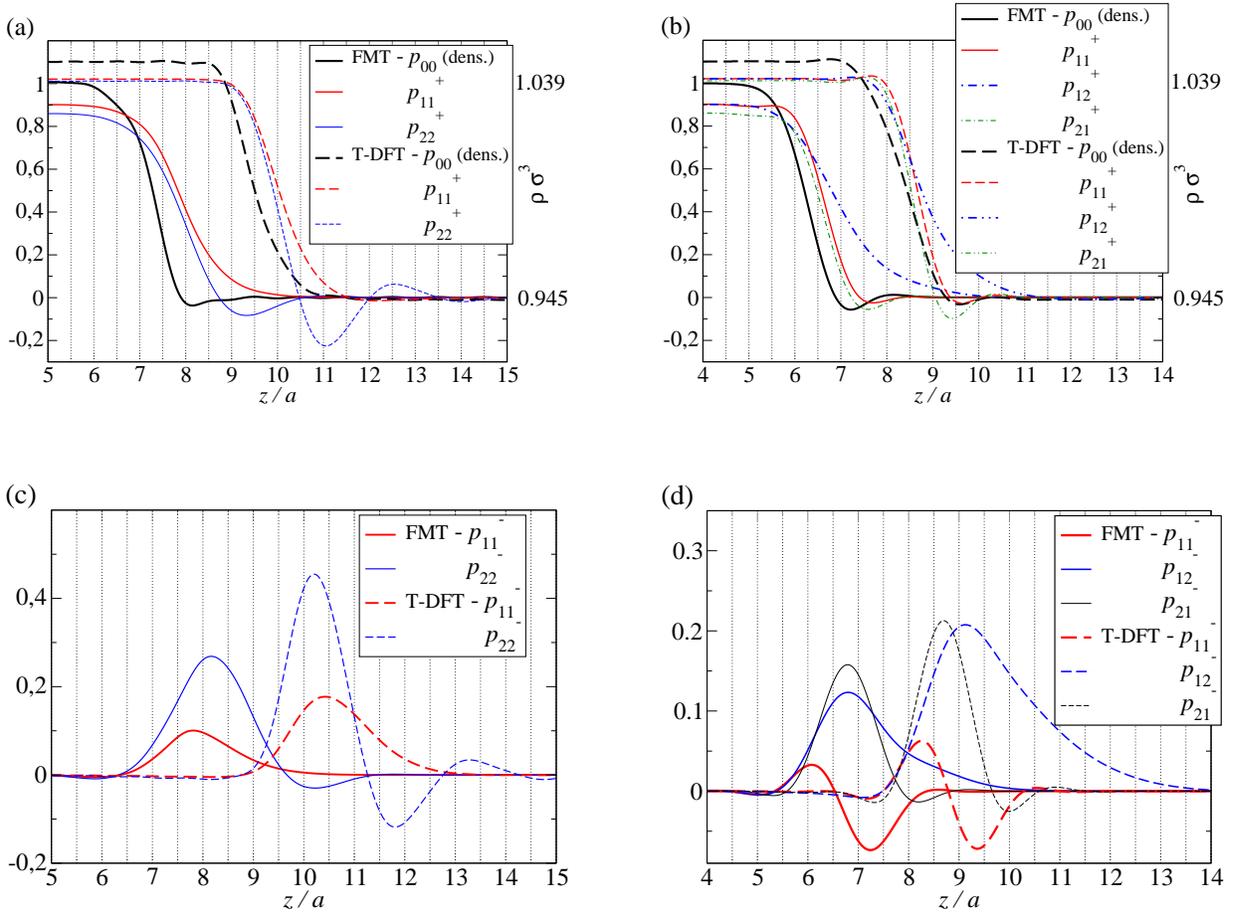

  \epsfig{file=fig6a.eps, width=7.5cm} \hspace{1cm}
  \epsfig{file=fig6b.eps, width=7.5cm} \\[10mm]
  \epsfig{file=fig6c.eps, width=7.5cm} \hspace{1cm}
  \epsfig{file=fig6d.eps, width=7.5cm}
  \caption[]{ Comparison between mode profiles from FMT and Taylor--expanded DFT (T--DFT, using
     the PY direct correlation function and $\rho_0 \sigma^3= 0.94$). Note that
     the interface position for T--DFT is shifted by about $2a$ compared to the interface
     position for FMT to enhance readability.
     Real part of leading density modes for (a) the [100] interface  and (b) the [111] interface. The density mode $p_{00}(z)$ has been rescaled and shifted,
     see tick labels at right $y$--axis.
     Imaginary part\tb{ s} of leading density modes are displayed in (c) for the [100]--interface  and in (d) for the [111]--interface.  }
  \label{fig:modes_fmt_tdft}
\end{figure}

The technique to perform the mode extraction from a full 3d solution $\psi(x,y,z)$ of a system
with a solid--liquid interface is described in Ref.~\cite{Oet12}. There, the density mode properties
for the FMT solutions of the hard--sphere interfaces in different orientations have been discussed in
detail. Some of these properties can be summarized as follows
\begin{enumerate}
 \item a separation of about one cubic unit cell length $a \approx 1.6$ $\sigma$
       between  the interface location as determined by the average density
       and the interface location as determined by the leading crystallinity mode ($p_{1n}(z)$)
 \item a small density depletion zone just in front of the bulk crystal (dip in profile $p_{00}(z)$)
 \item strongly non-monoton\tb{ ic} mode profiles also for next--to--leading modes, especially for
       $p_{2n}(z)$
 \item kink position for higher modes $p_{mn}(z)$ shifts towards the bulk crystal for increasing $m$
\end{enumerate}

\subsubsection{Mode profiles from T--DFT and PFC in comparison with FMT}

First, we compare the leading density modes of the [100] and [111] interfaces for the FMT solutions and the
solutions from Taylor--expanded DFT (T--DFT), see Fig.~\ref{fig:modes_fmt_tdft}. The meaning of the different
modes is seen best from the associated RLV. As a basis for the RLV, we choose vectors in direction of
the cartesian axes with lengths $2\pi/a_x$, $2\pi/a_y$ and $2\pi/a_z$, respectively. The quantities
$a_{x[y,z]}$ are the side lengths of the minimal cuboid fcc unit cells fitting the desired
interface orientation in $z$--direction. For a graphical representation, we refer to Ref.~\cite{Oet12} (Fig.~2).
In Fig.~\ref{fig:modes_fmt_tdft}, we illustrate the leading mode $p_{11}(z)$ for the [100] orientation (corresponding
to $\vect K = (1,1,1)$, the direction of close--packed planes) and the next--to--leading mode $p_{22}(z)$ (corresponding to
$\vect K =(0,0,2)$, leading oscillation of lateral density average).  For the [111] orientation, the leading mode
splits into $p_{11} \leftrightarrow \vect K =(0,2,1)$ and $p_{12} \leftrightarrow \vect K =(0,0,3)$. Both
RLV correspond to directions of close packed planes, but the mode $p_{11}^+(z)$ clearly differs from
$p_{12}^+(z)$. Only the latter has a monotonic shape  \tb{ as} expected for a ``leading--order" interface profile, similar
to $p_{11}^+(z)$ of the [100] orientation. For the next--to--leading mode, we have
$p_{21} \leftrightarrow \vect K =(1,1,2)$.
The FMT results for the real parts of these leading modes display already all  properties 1--4 listed above.
We further remark that the modes from T--DFT compare fairly well on a  semi--quantitative level.
The density depletion zone is missed and the phase shifts are more pronounced. For both, FMT and T--DFT, the
mode expansion converges slowly as seen by the plateau values of the modes in the crystalline bulk. This
is a consequence of the narrow density peaks in the bulk crystal.

\begin{figure}
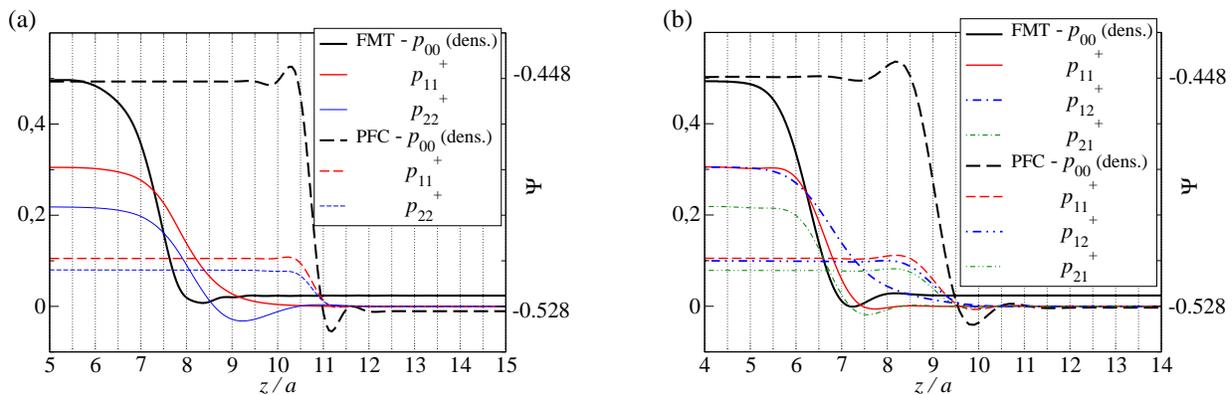

  \epsfig{file=fig7a.eps, width=7.5cm} \hspace{1cm}
  \epsfig{file=fig7b.eps, width=7.5cm}
  \caption[]{ Comparison between order parameter
   mode profiles from FMT and PFC for the [100] interface in (a) and
   the [111] interface in (b) for the PFC parameter $\epsilon=0.5$ (see Fig.~\ref{fig:phasediag_pfc}).
   The order parameter resulting from FMT is calculated from the 3d FMT density profile by
   $\Psi(\vect r) = \sqrt{g/(\rho_0^2 c_4 q_0^4)}\, [\bar\rho(\vect r)/\rho_0 - 1 - b/(3g)]$
   (see Eqs.~(\ref{eq:pfctrafo}) and (\ref{eq:rhobar}), $q_0^2=c_2/(2c_4)$). We choose
   the reference density $\rho_0 \sigma^3=0.94$. The remaining parameters are given in
   Table \ref{tab:pfcpars}. \tb{To enhance readability}  the ``density" mode $p_{00}(z)$ is rescaled and shifted(see tick labels at right $y$--axis) and that the interface position for PFC is shifted by about
     $3a$ compared to the interface
     position for FMT.   }
  \label{fig:modes_fmt_pfc}
\end{figure}

Next we compare the interface mode profiles for the order parameter $\Psi(\vect r)$ from PFC on the
one hand and from FMT on the other hand. Here, the FMT order parameter follows from
the smeared density $\bar\rho(\vect r)$ (Eq.~(\ref{eq:rhobar})) which is rescaled and shifted according
to the PFC transformation in Eq.~(\ref{eq:pfctrafo}). For the PFC parameter, $\epsilon=0.5$ and
a reference density of $\rho_0 \sigma^3=0.94$, the results for the real part of the leading modes
is shown in Fig.~\ref{fig:modes_fmt_pfc} (part (a) for the [100] interface and part (b) for the [111]
interface). First one notes that the absolute magnitude of the PFC modes is smaller by a factor of about
3 compared to the FMT modes. This is a consequence of the different widths of the order parameter peaks
in the crystal bulk: they are more narrow in FMT (see Fig.~\ref{fig:psi_comparison}) and consequently their Fourier
amplitudes are larger. Secondly, except for the ``density" depletion, the mode features identified in FMT
are not present in PFC. This is attributed to the simple free energy of the
PFC model. Our mode results further illustrate that the specificities of the layered hard sphere
packing can not be captured by PFC. Another interesting observation is that the interface width $w$
of the [100] interface is about $0.24\,a$ for PFC and $0.86\, a$ for FMT if the leading mode
$p_{11}(z)$ is fitted to the simple tanh--profile $1-\tanh[(z-z_0)/w]$. For
the [111] orientation we find widths of $0.47\,a$ (PFC) and $1.05\,a$ (FMT) from a fit to the
leading mode $p_{12}(z)$. In conclusion, we observe that the width of the interface is much  smaller in PFC and it varies considerably with the orientation of the interface.

\subsubsection{Mode profiles from simulations in comparison with FMT}

\tb{ We have carried out Molecular Dynamics simulations of the [100] interface 
in order to give a comparison to FMT data as well as to demonstrate the 
applicability of the mode expansion technique to simulation data.
The simulations were carried out in the $NVT$--ensemble at coexistence in cuboid boxes of cross--sectional area of 5 unit cells $\times$ 5 unit cells
($L_x\times L_y=7.84\sigma\times 7.84\sigma$)
and a length of $L_z \approx 205 \sigma$ with the crystal occupying about 60\% of the box volume and placed in the middle of it.
We have recorded the laterally averaged density profile
\bea
  \rho_{\rm av}(z) = \frac{1}{L_x L_y} \int_0^{L_x} {\rm d}x \int_0^{L_y} 
      {\rm d}y\, \rho(x,y,z)
  \label{eq:rhoav}
\eea
with a resolution of 64 points per unit cell as a time average over different 
time intervals $T_{\rm av}$.}

\tb{
From $\rho_{\rm av}(z)$ one can extract mode profiles $p_{mn}(z)$ for which 
the lateral components of the associated reciprocal lattice vector are zero: 
$(K_x)_{mn}=(K_y)_{mn} =0$. In particular we focused on the average density 
mode $p_{00}(z)$ and the first two modes appearing for the lateral
density average $p_{22}(z)$ ($(\vect K)_{22}=(0,0,2)$) and  
$p_{62}(z)$ ($(\vect K)_{62}=(0,0,4)$).
}

\tb{
Due to the periodic boundaries of the simulation box, global centre of 
mass motion of the system does not cost any free energy, hence the system 
diffuses freely.
When taking the time average, we have made no attempt to correct for 
this motion, as it was negligibly small on the time-scale of $T_{\rm av}$.  
Furthermore we have not corrected for
the zero mode of the capillary waves at the interfaces. The zero 
mode corresponds to a shift $\Delta z$
of the average interfacial position, which is caused by fluctuations in the 
overall amount of crystalline material. For an infinite system, this zero 
mode would not incur a free energy penalty, either.
For a finite system adsorption/desorption of crystalline layers results 
in a density change in the
surrounding liquid reservoir. This is associated with a free energy cost, 
which we estimate in quadratic approximation to be
\bea
 \Delta F = \frac{2 L_x L_y}{L_z}  (\rho_{\rm cr}-\rho_{\rm fl})^2 \mu'_{\rm fl}(\rho_{\rm fl})\;(\Delta z)^2\:,
\eea
where $\mu'_{\rm fl}(\rho)$ is the density derivative of the fluid chemical 
potential.
For our system the broadening of the interface due to the zero mode 
contribution is rather small,
$\langle (\Delta z)^2 \rangle < 1\;\sigma^{\tm{2}}$.
}

\tb{
Besides the zero mode, there are also capillary waves with a finite 
wavelength. It is impossible
to disentangle the contribution of capillary interfacial broadening from 
the width of a hypothetical ``intrinsic" density profile which one would like
to relate to the profile from density functional theory. However, by studying 
density averages for different time
intervals, we will obtain some qualitative insight regarding the contributation that capillary waves make to the density modes.
}

\begin{figure}
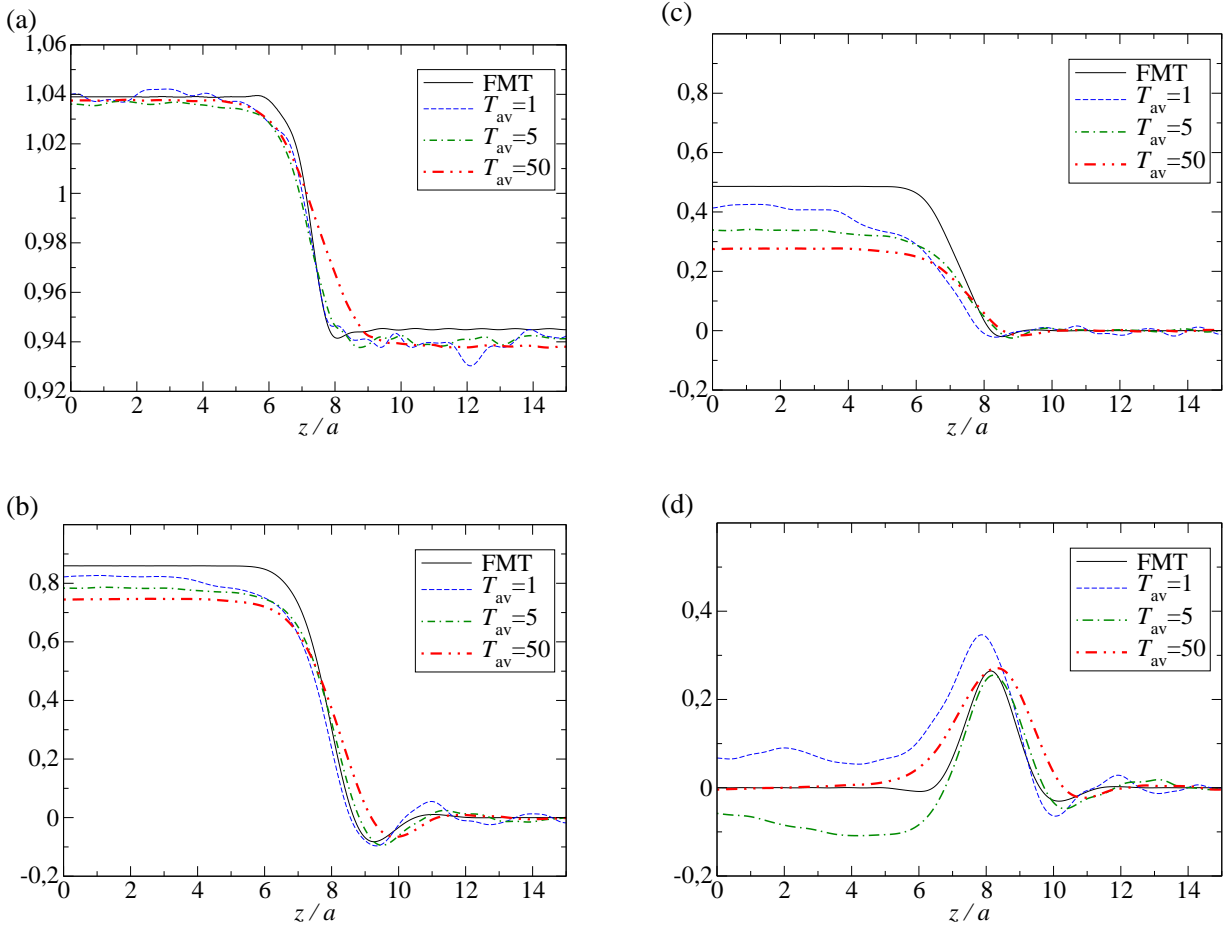

  \epsfig{file=fig8a.eps, width=7.5cm} \hspace{1cm}
  \epsfig{file=fig8c.eps, width=7.5cm} \\[0.7cm]
  \epsfig{file=fig8b.eps, width=7.5cm} \hspace{1cm}
  \epsfig{file=fig8d.eps, width=7.5cm}
  \caption[]{ Leading modes extracted from simulated, laterally averaged density profiles in comparison with FMT results.
 (a) average density mode $p_{00}(z)$, (b) real part of $p_{22}(z)$ ($(\vect K)_{22}=(0,0,2)$), (c)
 real part of $p_{62}(z)$ ($(\vect K)_{22}=(0,0,4)$)  and (d) imaginary part of $p_{22}(z)$.    }
  \label{fig:modes_sim}
\end{figure}

\tb{
In Fig.~\ref{fig:modes_sim} we show the mode profiles extracted from the 
simulation data for averaging times
of 1, 5 and 50 self--diffusion times $\tau$ as well as the FMT counterparts. 
(We give $T_{\rm av}$ in units of the characteristic self--diffusion 
time, which it takes a particle in
the coexisting liquid to diffuse over a distance of $\sigma$.) The width of 
the average density
(Fig.~\ref{fig:modes_sim}(a)) compares well with the FMT profile for $T_{\rm av}=1$ and 5 but shows a significant broadening
for $T_{\rm av}=50$. This we largely ascribe to finite--wavelength capillary waves which are sampled better at longer times.
The broadening effect on the mode width is less pronounced for the crystallinity modes (Fig.~\ref{fig:modes_sim}(b)--(d)). A strong effect
of the sampling time is visible on the plateau value in the crystalline part of the real parts of mode $p_{22}$ (Fig.~\ref{fig:modes_sim}(b)) and
of mode  $p_{62}$ (Fig.~\ref{fig:modes_sim}(c)). This reflects the broadening of the lattice site density peaks due to diffusion
of the crystal as a whole. Apart from that the behavior of the modes in the interface region $z/a=7\dots 11$ compares very well with
the FMT results. In particular the good agreement for the imaginary part of $p_{22}$ signifies that the wavelength shift of the
density oscillations across the interface is captured correctly by FMT.
}

\section{Summary and conclusion}

\label{sec:summary}

\tb{ We have studied the crystal--liquid interface in the hard sphere system 
by means of theoretical description on three approximative levels, all of  
which are based on classical density func\tm{t}ional theory (DFT):}
\begin{enumerate}
  \item fundamental measure theory (FMT), the currently most accurate theoretical framework for hard spheres,
  \item second--order Taylor--expanded DFT which truncates density fluctuations with respect to a reference density
      beyond second order,
  \item a phase field crystal (PFC) model which formally can be viewed as local expansion of the second-order Taylor-expanded DFT in density fluctuations
       and in gradients thereof to fourth order.
\end{enumerate}
Coexistence properties, surface tensions and interfacial profiles for three interface orientations
have been determined in all approximations
by free minimization of the associated density functionals. FMT provides us with benchmark results against
which the other approximations can be measured.
For the [100] interface, we have confirmed that also the interfacial density mode profiles obtained from FMT show good agreement with
corresponding results of Molecular Dynamics simulations. Thereby we have demonstrated the applicability of the
mode expansion technique to simulation data.

 An important difference between FMT on the one hand, \tb{ and} Taylor--expanded
DFT and PFC on the other hand is that the packing constraints for hard 
particles are incorporated very well only in the former. \tb{ In consequence, 
the small values for the relative vacancy concentration for 
equilibrium crystals and the relatively small surface tension anisotropy are predicted correctly}. Vacancy concentrations in
Taylor--expanded DFT and PFC are off by orders of magnitude. 
The surface tension anisotropy is too large
by a factor of two in Taylor--expanded DFT and much too large in PFC. In Taylor--expanded DFT, we find a surprising
sensitivity of coexistence properties to details of the direct correlation function. This  results in a strong sensitivity
of the average surface tension with respect to the choice of the direct correlation function.

We have discussed in detail the problem of parameter fixing in PFC. It turns out that the identification of the PFC
order parameter with a rescaled and shifted smeared density is a suitable working recipe. The structure factor and coexistence free energies
can be fitted very well. The order parameter distribution in the bulk crystal compares to FMT well only on a qualitative level,
but with regard to the surface tension and the interfacial structure there are actually big discrepancies. Our major conclusion here
is that the \tb{ simple PFC variant  considered here} is too generic and needs to be specifically modified in order to incorporate the hard--sphere like
interface structure of fcc materials, (see Ref.~\cite{Wu10} for an approach in this direction).

The sensitivity of results in Taylor--expanded DFT to the choice of the direct correlation function provides a hint that
 the specific functional form of the latter should be fitted to obtain proper coexistence. An attempt to match precise results
for the direct correlation function (obtained by other means) is perhaps of little practical use. We further discuss that, at least for the hard sphere system,
the status of Taylor--expanded DFT as the reference microscopic DFT for the PFC model is not justified.

For future work on statics and dynamics of crystalline interfaces, grain boundaries and defects, it  appears to be beneficial
to pursue both, microscopically precise calculations using FMT and more coarse--grained investigations using the PFC model.
Time and length scales accessible with FMT will be smaller than in PFC, but the PFC model can be better gauged in this way.

{\bf Acknowledgment:} The authors thank the DFG (German Research Foundation) for support through
the Priority Program SPP 1296, grants SCHI 853/2-2, OE 285/1-3 and NE 822/6-3). \tm{SD acknowledges the National Research Fund, Luxembourg co-funded under the Marie Curie Actions of the European Commission (FP7-COFUND) and the HPC facility of the University of Luxembourg for computing resources.}

\begin{appendix}

\section{Free minimization of the crystal--fluid interface in DFT}
\label{app:dftmin}

The free minimization of the crystal--fluid interface was achieved by an iterative solution of the Euler--Lagrange equation
\bea
  \label{eq:ele}
  \rho(\vect r) = \exp\left( \beta\mu - \frac{\delta (\beta{\cal F}^{\rm ex})}{\delta\rho(\vect r)} \right) = K[\rho]
\eea
where the excess free energy ${\cal F}^{\rm ex}$ is given by Eqs.~(\ref{eq:fex_fmt}) and (\ref{eq:phi_hs}) in the case of FMT and by Eq.~(\ref{eq:f_hnc}) in the
case of T--DFT, supplemented with the two choices for the direct correlation function (Eq.~(\ref{eq:c2power})), given by the coefficients
in Eqs.~(\ref{eq:c2tar}) and (\ref{eq:c2wbII}).
The density $\rho(\vect r)$ is discretized in a cuboid volume with edge lengths $L_x$, $L_y$
and $L_z$ which contains the fluid in the middle ($z\sim L_z/2$) and the crystal phase at the boundaries
($z \sim 0$ and $z \sim L_z$) such that $\rho(x,y,z)=\rho(x,y,L_z-z)$. $L_{x[y]}$ are given by the edge lengths in $x[y]$--direction, $a_{x,[y]}$,
of the smallest cuboid unit cell of the crystal
which has the desired orientation in $z$--direction. The crystal cuboid unit cells
are shown in Fig. 2 of Ref.~\cite{Oet12} for the [100], [110], and [111]
orientations. We chose $L_z=32 a_z$ for the [100] and [110] orientations and $L_z=16 a_z$ for the [111] orientation. The equidistant discretization was
usually 64 points per unit cell length $a_{x[y][z]}$, but was increased to 128 points per unit cell length $a_{y[z]}$ for the [111] case.
All convolution integrals appearing in $\delta {\cal F}^{\rm ex}/\delta\rho$ where calculated using 3d Fast Fourier transforms.

As a first step for initialization, bulk crystal density profiles have been determined for the coexisting state. To do so, we determined the minimal
free energy per particle of the bulk crystal in the cuboid unit cell
at a fixed bulk density $\rho_{\rm b}$ by solving Eq.~(\ref{eq:ele}) and also minimizing with respect to the unit cell length $a$ (corresponding to a
minimization with respect to the vacancy concentration \cite{Oet10}). The coexistence densities $\rho_{\rm cr}$, $\rho_{\rm fl}$
and the associated chemical potential $\mu=\mu_{\rm coex}$
were determined using the Maxwell construction.
The cuboid volume was filled with copies of the bulk crystal unit cell, defining a density profile $\rho_c(\vect r)$. The initial interface profile
$\rho_0(\vect r)$ was generated in the following way:
\bea
  \bar\rho(z)  & = &  \rho_{\rm fl} + (\rho_{\rm cr}-\rho_{\rm fl}) \; p(z,z_{0,1},w_1)\;, \\
  \rho_0(x,y,z) & = & \bar\rho(z) + (\rho_c(x,y,z) - \bar\rho(z)) \; p(z,z_{0,2}, w_2)\;, \\
  p(z,z_0,w)  & = & \frac{1}{2} \left( 2 - \tanh\left[ \frac{z-z_0}{w} \right] - \tanh\left[ \frac{L_z- z-z_0}{w} \right]   \right) \:.
\eea
The advantage of the above prescription is in the separation of the interfacial kink of the average density $\bar\rho$ from the kink of the
density oscillations through different choices for $z_{0,1}$ and $z_{0,2}$. This \tm{is} useful to ensure a smooth start into the iterations in the case of FMT,
for T--DFT it is not that important.

Iteration was done using a combination of Picard steps with variabe mixing and DIIS steps (discrete inversion in iterative subspace) \cite{Pul80}.
The Picard steps were performed according to
\bea
  \rho_{i+1}(x,y,z) & = & \alpha(z)\; K[\rho_i(x,y,z)] + (1-\alpha(z))\; \rho_i(x,y,z) \;, \\
   \alpha(z) & = & \alpha_{\rm min} + (\alpha_{\rm max}-\alpha_{\rm min}) p'(z,z_0,w) \;, \\
   p'(z,z_0,w) & = &  2 - \tanh^2\left[ \frac{z-z_0}{w} \right] - \tanh^2\left[ \frac{L_z- z-z_0}{w} \right]
\eea
The mixing function $\alpha(z)$ ensures that there are substantial changes within one iteration step only in the interfacial region, since
we chose $\alpha_{\rm min} \sim 10^{-5}$ and $\alpha_{\rm max} \sim 0.001...0.01$. Choosing $\alpha(z) = \mbox{const.}$ is not practical since
the constant would be limited to values below $10^{-4}$, otherwise the iteration fails due to instabilities  in the bulk crystalline region.
The DIIS steps were performed using between $n_{\rm DIIS}=7 ... 10$ previous profiles.

A typical FMT run consisted of an initial Picard sequence with about 50 steps and $\alpha_{\rm max}=0.01$. Then, we alternated between Picard
sequences of a minimum of 10 steps and  one DIIS step (which needs another $n_{\rm DIIS}$ Picard initialization steps).
We decreased $\alpha_{\rm max}$ after each switch from DIIS to Picard by a factor 1.5 until
$\alpha_{\rm max}=0.001$ was reached. Also, we varied the maximum position of the mixing function $\alpha(z)$ by choosing $z_0$ randomly in
a certain interval (located in the interface region) with a width of about 2 $\sigma$ after each switch from DIIS to Picard. This was done to overcome
being trapped at intermediate profiles where the surface tension
\bea
   \gamma = \frac{1}{2a_x a_y}\int_0^{a_x}dx \int_0^{a_y}dy \int_0^{L_z} dz
   \left( \beta^{-1} \rho(\vect r)(\ln \rho(\vect r) -1) +  f^{\rm ex}[\rho(\vect r)] -\mu\rho(\vect r) + p_{\rm coex} \right)
\eea
hardly changes between iteration steps ($f^{\rm ex}$ is the excess free energy density and $p_{\rm coex}$ is the coexistence pressure).
The condition for switching from Picard to DIIS was that after the minimum of 10 steps the convergence parameter
\bea
  \epsilon_i =  \frac{1}{a_x a_y L_z} \int_0^{a_x}dx \int_0^{a_y}dy \int_0^{L_z} dz \left( K[\rho_i(\vect r)]-\rho_i(\vect r)   \right)^2
\eea
was decreasing between subsequent steps. If not, the Picard iterations were repeated with another 10 steps until that condition was met.
Otherwise DIIS might take one away from the equilibrium solution easily. The DIIS step usually resulted in a very noticeable change in $\gamma$ and
also in $\epsilon_i$ in the subsequent Picard steps. It was, however, not possible in general to perform a second DIIS step immediately after the
first one \tr{since the density profile obtained after this DIIS step lead to singularities in the
free energy (local packing fraction $n_3>1$)}. We stopped the run when $\epsilon_i \alt 10^{-3}$.

We emphasize that only through the  procedure outlined above  we were able to determine equilibrium profiles for FMT.
The standard method for solving DFT, simple Picard iterations
with possibly variable, but spatially constant mixing $\alpha$, simply fails. Also without DIIS we were not able to arrive at equilibrium profiles
within a reasonable time.

For T--DFT, the above procedure does not seem to be necessary but resulted in a very quick convergence.


\section{Minimization of the PFC free energy for the crystal--fluid interface}
\label{app:pfcmin_interface}

In PFC, we perform simulations with periodic boundary conditions in each direction, as we do in DFT. In the crystal phase, this implies that
a stress will be acting on the crystal unless
the dimensions $L_{x[y][z]}$ of the cuboid simulation box fit exactly multiples of the corresponding unit cell lengths of the equilibrium crystal. 
In order to avoid this stress, we use a simulation box which is minimizing the free energy of the crystal, i.e. we determine
the minimizing length of the cubic unit cell $a_{\rm min}:= 2\pi/q$ of the fcc crystal (given in dimensionless PFC coordinates, $\vect x = q_0 \vect r$).
For a given average order parameter $\bar\Psi_\text{cr}$, we apply Brent's method to find the box length which minimizes the free energy. 

A test for a single crystal cubic unit cell in [100] orientation (for $\epsilon=0.5$ at coexistence, $\bar\Psi_\text{cr}=-0.448336$) 
with numbers of points per direction $N=8$, 16, 32 and 64 has shown that numerical box effects disappear for cubes of 
edge length 16 and larger. 
The results for the reciprocal lattice parameters $q$ are $q^{(8)}=0.539898$, $q^{(16)}=0.539469$, $q^{(32)}=0.539476$ and $q^{(64)}=0.539468$.
It is interesting to compare these numbers to the corresponding numbers obtained by expanding the crystal order parameter in reciprocal
lattice vectors (see Sec.~\ref{subsec:mode_theory}) and cutting the expansion at a maximum number $n_{\rm sh}$ for the reciprocal lattice vector shells. 
We find for $n_{\rm sh}=4$, 6, 8 and 10 the values $q^{(4)}=0.53990$, $q^{(6)}=0.53989$, $q^{(8)}=0.53956$ and $q^{(10)}=0.53948$.
This demonstrates that  $n_{\rm sh}$ corresponds roughly to $N/2$ and that for precise numerical results the few--mode approximation is not quite sufficient.

In order to avoid numerical artifacts, we determine the minimal reciprocal lattice parameters $q$ separately for each orientation; we simulate 
one unit cell of the crystal with $N= 32$  for the  [100] and the [110] interface. 
The crystal unit cell in [111] orientation is simulated in a box with discretization $32\times64\times64$. The cuboid crystal unit cells for the different orientations
are the same as used in the DFT calculations  (see Fig. 2 of Ref.~\cite{Oet12}). 

For the initialization of simulations of the crystal--liquid interface, half of the simulation box is filled with a one mode approximation of the crystal,
in [100] orientation given by
\bea
  \Psi (x) &=& \bar\Psi_\text{cr} + A\cos(qx)\cos(qy)\cos(qz),
\eea
and the liquid part in the other half has the constant value at coexistence $\bar\Psi_\text{fl}$. 
The box length in the $z$--direction (perpendicular to the interface) is 32 crystal unit cell lengths  for the [100] and [110] orientation, resulting 
in a simulation box with a total number of points of $32\times32\times1024$. For the [111] interface, we use 16 unit cell lengths resulting in 
a box with a total number of points of  $32\times64\times1024$. The crystal resides in one half of the box, so that 
one interface\tm{s} is in the middle of the box and the other near the periodic boundary.

The PFC simulation evolves according to the dynamic equation (\ref{eq:pfc_dyn_eq}) until the system relaxes. 
As an indicator for the relaxation, we use the average deviation $\delta\mu$ of the local chemical potential $\mu(\vect x)= \delta F_{\rm PFC}/\delta \Psi(\vect x)$
from the coexistence value $\mu_{\rm coex}$ and stopped the computation when $\delta\mu \sim 10^{-4}$.

For the calculation of the dimensionless surface tension $\tilde\gamma$ we use the formula
\begin{align}
  2\tilde\gamma = \frac{1}{\Omega} \int {d^3}{x} \left[ f -\left( f_{\rm cr} \frac{\Psi-\bar\Psi_{\rm fl}}{\bar\Psi_{\rm cr}-\bar\Psi_{\rm fl}} - 
f_{\rm fl} \frac{\Psi-\bar\Psi_{\rm cr}}{\bar\Psi_{\rm cr}-\bar\Psi_{\rm fl}}\right) \right] \label{eq:1}
\end{align}
from \cite[Eq.~(50)]{Wu07}, where $f$ is the PFC free energy density ($f_{\rm cr}$ for the crystal
at coexistence and $f_{\rm fl}$ for the liquid at coexistence). $\Psi$ denotes the PFC order parameter with $\bar\Psi_{\rm cr}$ the order parameter average 
in the coexisting crystal and $\bar\Psi_{\rm fl}$ the corresponding average  in the coexisting liquid. $\Omega$ is the interface area (in dimensionless PFC units).
Upon reordering Eq.~\eqref{eq:1} we find:
\begin{align}
  2\tilde\gamma &= \frac{1}{\Omega}  \int d^3x \left[ f +\frac{ f_{\rm cr}\bar\Psi_{\rm fl}-f_{\rm fl}\bar\Psi_{\rm cr}}{\bar\Psi_{\rm cr}-\bar\Psi_{\rm fl}} - \frac{\Psi}{\bar\Psi_{\rm cr}-\bar\Psi_{\rm fl}}\left(f_{\rm cr}-f_{\rm fl}\right) \right]\\
  &= \frac{1}{\Omega}  \left( \int d^3x \left[ f \right] +\frac{ f_{\rm cr}\bar\Psi_{\rm fl}-f_{\rm fl}\bar\Psi_{\rm cr}}{\bar\Psi_{\rm cr}-\bar\Psi_{\rm fl}} - \frac{f_{\rm cr}-f_{\rm fl}}{\bar\Psi_{\rm cr}-\bar\Psi_{\rm fl}} \int d^3x \left[ \Psi \right] \right) \\
  \shortintertext{define $\bar{f}, \bar{\Psi}$ as volume averages of the free energy density and order parameter}
  &= \frac{1}{\Omega}  \int d^3x \left[ \bar{f}  +\frac{ f_{\rm cr}\bar\Psi_{\rm fl}-f_{\rm fl}\bar\Psi_{\rm cr}}{\bar\Psi_{\rm cr}-\bar\Psi_{\rm fl}} - \frac{f_{\rm cr}-f_{\rm fl}}{\bar\Psi_{\rm cr}-\bar\Psi_{\rm fl}} \bar\Psi \right] \\
  &= \frac{V}{\Omega}  \left[ \bar{f}  +\frac{ f_{\rm cr}\bar\Psi_{\rm fl}-f_{\rm fl}\bar\Psi_{\rm cr}}{\bar\Psi_{\rm cr}-\bar\Psi_{\rm fl}} - \frac{f_{\rm cr}-f_{\rm fl}}{\bar\Psi_{\rm cr}-\bar\Psi_{\rm fl}} \bar\Psi \right],
\end{align}
where $V=\int d^3x=L_x\cdot L_y \cdot L_z$ so $\frac{V}{\Omega}= L_z$, and we obtain
\begin{align}
  \tilde\gamma=\frac{L_z}{2}\left( \bar{f} + \frac{ f_{\rm cr}\bar\Psi_{\rm fl}-f_{\rm fl}\bar\Psi_{\rm cr} - \bar\Psi(f_{\rm cr}-f_{\rm fl})}{\bar\Psi_{\rm cr}-\bar\Psi_{\rm fl}}\right) . \label{eq:gamma}
\end{align}
Note that the factor $\frac{1}{2}$ is needed due to the presence of two interfaces in the simulation.

To calculate the surface tension with Eq.~\eqref{eq:gamma} we calculate $\bar{f}, \bar{\Psi}$ in the whole domain. $\bar\Psi_{\rm fl}$, $\bar\Psi_{\rm cr}$, $f_{\rm cr}$ and $f_{\rm fl}$ are calculated by convoluting $f$ and $\Psi$ with normalized Gaussians of sufficient width such that the resulting profile is locally constant. 
This is equivalent to peak to peak averaging of the $f$ and $\Psi$ profiles on the crystal side.

For $\epsilon=0.53$, results for the surface tension have been reported previously in Ref.~\cite{Ta11}.
We checked the convergence of the surface tension for different number of points $N$ per unit cell length. 
For the [100] orientation, the results are given in  Tab.~\ref{tab:gammaconvergence} and should be compared with $\gamma_{[100]}=0.0113$ from Ref.~\cite{Ta11}.
\begin{table}
  \begin{tabular}{rr} \hline \hline
    $N$ & $\tilde\gamma_{[100]}$\\\hline
    8  & 0.00913 \\
    16 & 0.01041 \\
    32 & 0.01041 \\
    64 & 0.01052 \\ \hline \hline
  \end{tabular}
  \caption{Surface tension $\tilde\gamma$ for $\epsilon=0.53$ in $[100]$-direction.}  \label{tab:gammaconvergence}
\end{table}
For the $[111]$-direction Ref.~\cite{Ta11} provides $\gamma_{[111]}=0.0082$, whereas our result is $\gamma_{[111]}=0.0079$ (using $N= 32$).
It is not clear which precise discretization was used in Ref.~\cite{Ta11} but we can conclude that typical discretizations of about 10..15 points
per unit cell used in the PFC community leave a residual error of about 5 per cent in the value of the surface tension.

\pagebreak

\end{appendix}

\end{document}